\documentclass[fleqn,usenatbib]{mnras}

\usepackage[T1]{fontenc}
\usepackage{ae,aecompl}

\usepackage{graphicx}	
\usepackage{amsmath}	
\usepackage{newtxtext,newtxmath}

\newcommand{\hb}{H$\beta$}
\newcommand{\ha}{H$\alpha$}

\newcommand{\hei}{He\,\textsc{i}}
\newcommand{\heii}{He\,\textsc{ii}}
\newcommand{\siii}{\textsc{[S\,iii]}}

\newcommand{\oii}{[O\,\textsc{ii}]}
\newcommand{\neiii}{[Ne\,\textsc{iii}]}
\newcommand{\sii}{\textsc{[S\,ii]}}

\newcommand{\cii}{[C\,\textsc{ii}]}
\newcommand{\cliii}{[Cl\,\textsc{iii}]}
\newcommand{\oi}{[O\,\textsc{i}]}

\newcommand{\niA}{\textsc{[N\,i]}}
\newcommand{\nii}{\textsc{[N\,ii]}}
\newcommand{\niii}{\textsc{[N\,iii]}}
\newcommand{\ariii}{[Ar\,\textsc{\,iii]}}
\newcommand{\oiii}{\textsc{[O\,iii]}}

\title[The planetary nebula NGC 3132 revisited]{The planetary nebula NGC 3132 revisited: high definition 3D photoionization model}

\author[Monteiro et al.]{
H. Monteiro$^{1,2}$\thanks{E-mail: hmonteiro@unifei.edu.br},
C. Mendes de Oliveira$^{3}$,
P. Amram$^{4}$,
L. Stanghellini$^{5}$,
R. Wesson$^{1}$,
\newauthor{K. Bouvis$^{6,8}$,
S. Akras$^{6}$,
M. Matsuura$^1$,
B. C. Quint$^{7}$}
\\
$^{1}$ School of Physics and Astronomy, Cardiff University, Queen's Buildings, The Parade, Cardiff CF24 3AA, UK\\
$^{2}$ Universidade Federal de Itajub\'a - UNIFEI\\
$^{3}$Departamento de Astronomia, Instituto de Astronomia, Geof\'isica e Ci\^encias Atmosf\'ericas da USP, Cidade \\ Universit\'aria, 05508-900, S\~ao Paulo, SP, Brazil\\
$^{4}$ Aix Marseille Universit\'e, CNRS, LAM (Laboratoire d'Astrophysique de Marseille), Marseille, France\\
$^{5}$ NSF's National Optical-Infrared Astronomy Research Laboratory, Tucson, AZ 85719\\
$^{6}$ Institute for Astronomy, Astrophysics, Space Applications and Remote Sensing, National Observatory of Athens, GR 15236 Penteli, Greece\\
$^{7}$Rubin Observatory Project Office, 950 N. Cherry Ave., Tucson, AZ 85719, USA\\
$^{8}$Department of Physics, University of Patras, Patras, 26504 Rio, Greece}

\date{Accepted XXX. Received YYY; in original form ZZZ}

\pubyear{2019}

\begin{document}
\label{firstpage}
\pagerange{\pageref{firstpage}--\pageref{lastpage}}
\maketitle

\begin{abstract}
We present a detailed 3D photoionization model of the planetary nebula NGC 3132, constrained by the latest observations. Using the MOCASSIN code, the model incorporates integrated and spatially resolved spectroscopy, velocity-resolved line profiles, emission line maps, and photometry, including recent high-quality data from MUSE (VLT) and JWST among others. Based on new data from the SAMFP instrument at SOAR, the three-dimensional density structure of the nebula was obtained by assuming homologous expansion of the surrounding nebular gas. The final fitted model successfully reproduces all key observational constraints available, particularly in terms of the detailed emission line integrated fluxes and ionization structures across different ionic stages. The results of the model show that the progenitor star had a mass of $(2.7 \pm 0.2)M_{\odot}$ and is surrounded by a He poor shell of dust and gas. The abundances of He, C, N, O, and S determined by the model show that the nebula has C/O=$(2.02 \pm 0.28)$ and N/O=$(0.39 \pm 0.38)$ consistent with the progenitor mass found.

\end{abstract}




\section{Introduction}
\label{sec:intro}

Planetary nebulae (PNe) are the ejected envelopes of intermediate-mass stars that have recently terminated their asymptotic giant branch stage of evolution. These objects are of great interest to astronomers because they are unique laboratories for studying the late stages of stellar evolution, and the physics of the interplay between the central star and the surrounding gas. 

One such object is the nebula NGC~3132 (PN G272.1+12.3), which due to its large size and brightness has been well studied throughout the years. Based on narrow-band emission line imaging data, which showed a clear elliptical morphology, various authors originally assumed a closed ellipsoidal geometry for the gas distribution of NGC~3132 \citep{Masson1990,Zhang1998, BDG90}. Based on high-resolution spectra of the \oiii~$\lambda$5007 emission line in five positions, \citet{SD86} also proposed a closed ellipsoidal shell model with velocity and density asymmetries to explain the observations. However, closed elliptical shell models were unable to reproduce the low central density observed by \citet{SD86} and \citet{Juguet1988} nor the observed asymmetric double-peaked velocity profiles in regions far from the centre of the nebula. \citet{Sahai1990} discussed and also ruled out the \citet{SD86} model based on the detected CO emission, which showed that the neutral gas exists in an equatorial torus. 

The availability of many spectroscopic observations \citep{AF64,K76,TP77} allowed more detailed modelling efforts. In particular, \citet{BDG90} using their own spectroscopic data calculated the first three-dimensional photoionisation model for NGC 3132. Their results based on the reproduction of the H${\beta}$ image and total flux indicated that a closed elliptical shell with varying density along all axes could explain the observations well.

In \citet{Monteiro2000} a three-dimensional photoionisation model was used to study the morpho-kinematic properties of NGC 3132 showing that although a closed shell was able to reproduce the H${\beta}$ image, it did not reproduce the density profile obtained from \sii$\lambda\lambda$6716,6731 or the double peaked asymmetrical structure of the velocity profiles in the outer regions based on the high-resolution data from \citet{SD86}.  However, an inclined bipolar diabolo-like shape was able to reproduce the observed images, velocity profiles, and high-resolution observations. These results were in reasonable agreement with the expansion velocities measured by \citet{MWF88} using \oiii~$\lambda$5007 and \oii~$\lambda\lambda$3726,3729 emission lines, obtained in slits along the major axis, where values of 14.7~km~s$^{-1}$ and  21~km~s$^{-1}$ respectively were found. The results were also able to reproduce the asymmetric and double-peaked profiles obtained by \citet{SD86}.

In the same way that models and some observations indicated discrepancies in the distribution of matter inferred for the nebula, the general properties of its central star were also considerably uncertain. The calculated values for the Zanstra \ion{He}{2} temperature of the ionising star of NGC~3132 ranged from 73 000 K \citep{Pacheco1986} to 110 000 K \citep{Pottasch1996}. The ionising star luminosity values available in the literature varied from 72 L$_{\odot}$ \citep{Mendez1978} to $\sim$ 125 L$_{\odot}$ \citep{Pottasch1984} with a recent value of 250 L$_{\odot}$, obtained from a one-dimensional photoionization model by \citet{MUSE3132}. 

Other properties were also reasonably established such as the observed H$\beta$ flux (on logarithmic scale) ranging from -10.49 $erg/cm^2/s^{-1}$ \citep{1971BAICz..22..103P} to -10.20 $erg/cm^2/s^{-1}$ \citep{1977A&A....54..435P}. The extinction has been measured by many authors \citep{1977A&A....54..435P,1978MNRAS.185..647M,1982AJ.....87..555F,Gathier1986}, with the color excess E(B-V) of the order of 0.1.
 
This interesting object continued to be studied in detail with more modern instruments. In \citet{Liu2001} the authors presented observations of far-infrared (IR) spectra obtained using the Long Wavelength Spectrometer (LWS) on the Infrared Space Observatory (ISO). The obtained spectra provided flux measurements for fine-structure lines emitted in the ionized regions ([N~{\sc ii}] 122$\micron$, [N~{\sc iii}] 57$\micron$, [O~{\sc iii}] 52$\micron$, and 88$\micron$) as well as lines from the photodissociation regions (PDRs) ([O~{\sc i}] 63$\micron$ and 146$\micron$, [C~{\sc ii}] 158$\micron$). These measurements were used to determine electron densities, ionic abundances, temperatures, and gas masses for both the ionised regions and the PDRs. For NGC 3132 the authors find the abundance ratio C/O=0.6, C/H=$6\times 10^{-4}$ and in the PDR a temperature of T=$220$K and a density of log($N_H$)=4.9~cm$^{-3}$.  

In \citet{Tsamis2003} the authors present deep optical spectrophotometry of 12 Galactic planetary nebulae (PNe) including NGC~3132 and three Magellanic Cloud PNe. They obtain mean spectra by uniformly scanning the long-slit of the spectrograph across the nebular surfaces. The authors also present observations from ultraviolet (UV) and infrared (IR) spectra acquired by space-based instruments such as the International Ultraviolet Explorer (IUE) and Infrared Space Observatory (ISO) satellites.  In \citet{Tsamis2004} the authors continue the study of the same sample of PNe now focusing on the analysis of the relative intensities of faint optical recombination lines (RLs) finding a significant abundance discrepancy factor (ADF) for many objects. They concluded that the main cause of the discrepancy is enhanced ORL emission from cold ionised gas located in hydrogen-deficient clumps inside the main body of the nebulae. For NGC 3132 they find an ADF of 4.0, 3.5 and 2.4 for C, N and O respectively.

Infrared images were obtained in \citet{Hora2004} as part of an imaging survey of planetary nebulae (PNs) using the Infrared Array Camera (IRAC). In NGC~3132, the 8.0$\micron$ emission is seen to extend further from the central star compared to the optical and other IRAC bands, possibly due to the presence of H$_2$ emission and the distribution of molecular material outside the ionized regions.

High signal-to-noise ratio spectroscopic observations using long-slit in the 3100 to 6900\AA~ range were also obtained by \citet{Krabbe2005} and emission-line intensities were presented in \citet{Krabbe2006}. The authors find similar abundances to \citet{Tsamis2003}, as well as a systematic spatial variation of electron density in NGC~3132. 

An atlas of Hubble Space Telescope (HST) images and ground-based, long-slit, narrow-band spectra centred on the 6584\AA~ line of \nii~and the 5007\AA~  line of \oiii~was presented by \citet{Hajian2007}. The spectra, which were obtained for a variety of slit positions across each target, were combined with a prolate ellipsoidal model to obtain information on the velocity field. For NGC 3132, they find an equatorial expansion velocity $v_0$ of 33~$\rm{km/s}$ for the \nii~line and 14~$\rm{km/s}$ for the \oiii~line. We also note that additional HST / WFPC2 images for \ha, \oi~6300\AA, \oiii~5007\AA,  \nii~6583\AA, and \sii~6717,6731\AA~are available at the Mikulski Archive for Space Telescopes (MAST) \footnote{\url{https://archive.stsci.edu/}}.

In the Chandra Planetary Nebula X-Ray Survey (ChanPlaNS) conducted by \citet{Kastner2012}, NGC 3132 did not show any X-ray detection. The study revealed that most of the X-ray undetected planetary nebulae (PNe) possess high molecular content and display distinct bipolar or ring-like morphologies. According to the authors, the absence of X-ray emission from the central stars in these PNe could be attributed to the presence of magnetically inactive central stars or merged companions. The non detection may also indicate that these particular PNe undergo a more rapid evolutionary process compared to others.

Infrared spectra from Spitzer/IRS were also analysed by \citet{Delgado2014} to search for crystalline silicates, polycyclic aromatic hydrocarbons (PAHs), and other dust features. They indicated that only crystalline silicates are detected in NGC 3132. The authors determined the iron abundance for the object, finding  a range of [-3.6, -3.2] for $\Delta[Fe/O]$,  indicating significant iron depletion of dust grains. They also used literature values to recalculate abundances and determined $log[C/O]=-0.47$ from collisionally excited lines (CELs) and $log[C/O]=-0.11$ from RLs. Spitzer spectroscopy observations were also presented by \citet{Mata2016}, where they report the detection in NGC 3132 of mid-IR ionic lines of [Ar III], [S IV], [Ne II] and [Ne V], as well as polycyclic aromatic hydrocarbon features and other $H_2$ lines. Analysis of the population distribution of the $H_2$ molecules revealed an excitation temperature of $T_{ex}(rot)=900$ K.

The distances determined before Gaia were uncertain and ranged from 0.51 kpc \citep{Gathier1986,Pottasch1996} to 1.63 kpc \citep{TP77} with other works finding values in that range \citep{Stanghellini2008,Frew2016}. The most recent distance estimate for the nebula is from \citet{2022NatAs...6.1421D}, based on the distance of the brighter A-type companion of the central ionising star. The bright A star has a Gaia DR3 geometric distance of 754 pc (with uncertainties of $+18$pc and $-15$pc) determined by \citet{Bailer-Jones2021} and the same radial velocity of the nebula, while the faint central star's Gaia DR3 distance of 2,125pc (with uncertainties of $+559$pc and $-1,464$pc) has poor astrometric quality, likely due to the nearby bright A star.

 One of the most detailed spectroscopic studies of NGC\,3132 is presented in \citet{MUSE3132}, where the authors have obtained detailed 2D spectroscopic data acquired during the MUSE instrument commissioning on the ESO Very Large Telescope. They show that the nebula presents a complex reddening structure with high values (c(\hb)$\sim0.4$) at the rim. They also find that density maps are compatible with an inner high-ionisation plasma at moderate high density ($\sim$1000~cm$^{-3}$), while the low-ionisation plasma presents a structure in density, which peaks at the rim with values $\sim$700~cm$^{-3}$. Median $T_e$ using different diagnostics decreases according to the sequence \nii,\sii $\rightarrow$ \siii  $\rightarrow$ \oi $\rightarrow$ \hei $\rightarrow$ Paschen Jump. The range of temperatures covered by recombination lines is much larger than those obtained from collisionally excited lines (CELs), with large spatial variations within the nebula. They determined a median helium abundance He/H= 0.124, with slightly higher values at the rim and outer shell. Their results show that velocity maps support a geometry for the nebula similar to the diabolo-like model proposed in \citet{Monteiro2000}, but oriented with its major axis roughly at P.A.$\sim-22^\circ$. 

 \citet{MUSE3132} also identified two low-surface brightness arc-like structures towards the northern and southern tips of the nebula, with high extinction, high helium abundance, and strong low-ionisation emission lines. They are spatially coincident with some extended low-surface brightness mid-IR emission. They suggest that these characteristics are compatible with being the consequence of precessing jets caused by the binary star system.

Observational data from the James Webb Space Telescope (JWST) Early Release were discussed in detail in \citet{2022NatAs...6.1421D}. The authors find a structured, extended hydrogen halo surrounding an ionized central bubble which exhibits spiral formations, likely due to a low-mass companion orbiting the central star. The observations also uncovered a mid-infrared excess near the central star, suggesting the presence of a dusty disk, likely resulting from an interaction with a closer companion. The JWST images enabled the development of a model detailing the illumination, ionization, and hydrodynamics of the molecular halo, highlighting the complexities of the stellar outflows. New data on the A-type visual companion allowed a precise estimation of the progenitor star mass of $2.86 \pm 0.06$ M$_{\odot}$. These findings, in particular, about the central dust disk, were also corroborated by the results of \citet{2023ApJ...943..110S}.

The latest data obtained for NGC 3132 is from \citet{2024ApJ...965...21K}, where the authors present Submillimeter Array (SMA) observations mapping the $^{12}$CO $J = 2 \rightarrow 1$, $^{13}$CO $J = 2 \rightarrow 1$, and CN $N = 2 \rightarrow 1$ emission at 5" resolution. The data show that a molecule-rich, bright ring is an expanding structure rather than a limb-brightened shell, indicating a bipolar planetary nebula viewed nearly pole-on. In addition, a second expanding molecular ring, oriented nearly orthogonally to the main ring, is identified. The authors argue that the two-ring morphology likely originates from the disruption of an ellipsoidal molecular envelope by misaligned, fast collimated outflows, possibly driven by interactions with one or more companions during the late evolutionary stages of the progenitor star.

In this context, the present paper presents a detailed three-dimensional photoionization model for NGC 3132 constrained by the most up-to-date observational data available. We use a density structure derived from new spatially resolved high-resolution data obtained with the scanning Fabry-Perot interferometer (SAM-FP) instrument attached to the SOAR telescope, revealing details of the velocity structure of the surrounding nebular gas. This structure is then used to obtain a high definition photoionization model for NGC 3132. In Sec. \ref{sec:obs} we present previously unpublished SAM-FP observational data used to derive the density structure as well as additional IUE data remeasured to account for aperture effects. These observations, together with the other data discussed previously, comprise the constraints adopted for the modelling effort. In Sec. \ref{sec:3Dmodel} we describe the details of how the gas and dust density structures are determined and how the central ionizing source was defined. In that section, we also describe the implementation of an optimization procedure used to obtain the best fit for abundances of He, C, N, O and S, allowing us to set reliable confidence intervals in the derived values. In Sec. \ref{sec:results} we present and discuss the results of the best fit model and in Sec. \ref{sec:conclusion} we give our final conclusions.

\section{Observational data for NGC~3132} 
\label{sec:obs} 

To obtain a detailed and accurate three-dimensional model of NGC 3132 we require a wide range of observational constraints, in particular those that can provide spatially resolved information. In that sense, we have collected diverse observational material from the literature ranging from integrated line fluxes and photometry to spatially resolved spectroscopic data. Many of the constraints that will be used in this work have already been presented and discussed in Sec. \ref{sec:intro}. In what follows, we focus on the new data obtained with the SOAR-FP instrument, which was used to define the three-dimensional gas and dust structure used to model the nebula. We also discuss emission line fluxes from IUE observations which were remeasured to be used as constraints.

\subsection{Observations from the SAM-FP}
\label{sec:samfp}

Two observations of NGC~3132 were taken on UT date April 1, 2017, with the [N~{\sc ii}] filter (SAMI 6584-20) and on UT date April 3, 2017, in H$\alpha$ (filter BTFI 6569-20), with the high-resolution scanning Fabry-Perot interferometer SAM-FP, mounted on the SOAR telescope adaptive module (see \cite{MendesOliveira2017}, for details on the Fabry-perot used, as well as the work by \cite{Derlopa2024}). The Fabry-Perot system utilised in this study was the ICOS ET-65 from the Fabry-Perot Company, with an interference order of 609 at 6562.78 angstroms. The free spectral range of the interferometer (492~km~s$^{-1}$) was covered in 38 and 43 scanning steps, respectively. Each scanning step corresponded to 0.28 angstroms (equivalent to 12.8~km~s$^{-1}$). The observations were captured using an e2v CCD detector with dimensions of 4096 $\times$ 4112 pixels. The total field of view covered an area of 180 arcseconds $\times$ 180 arcseconds, with a pixel size of 0.18 arcseconds. The readout noise, when unbinned, was measured to be 3.8 electrons, and the gain was estimated at 2.1 electrons per analog-to-digital unit. 

\begin{figure*}
	\includegraphics[width=\columnwidth]{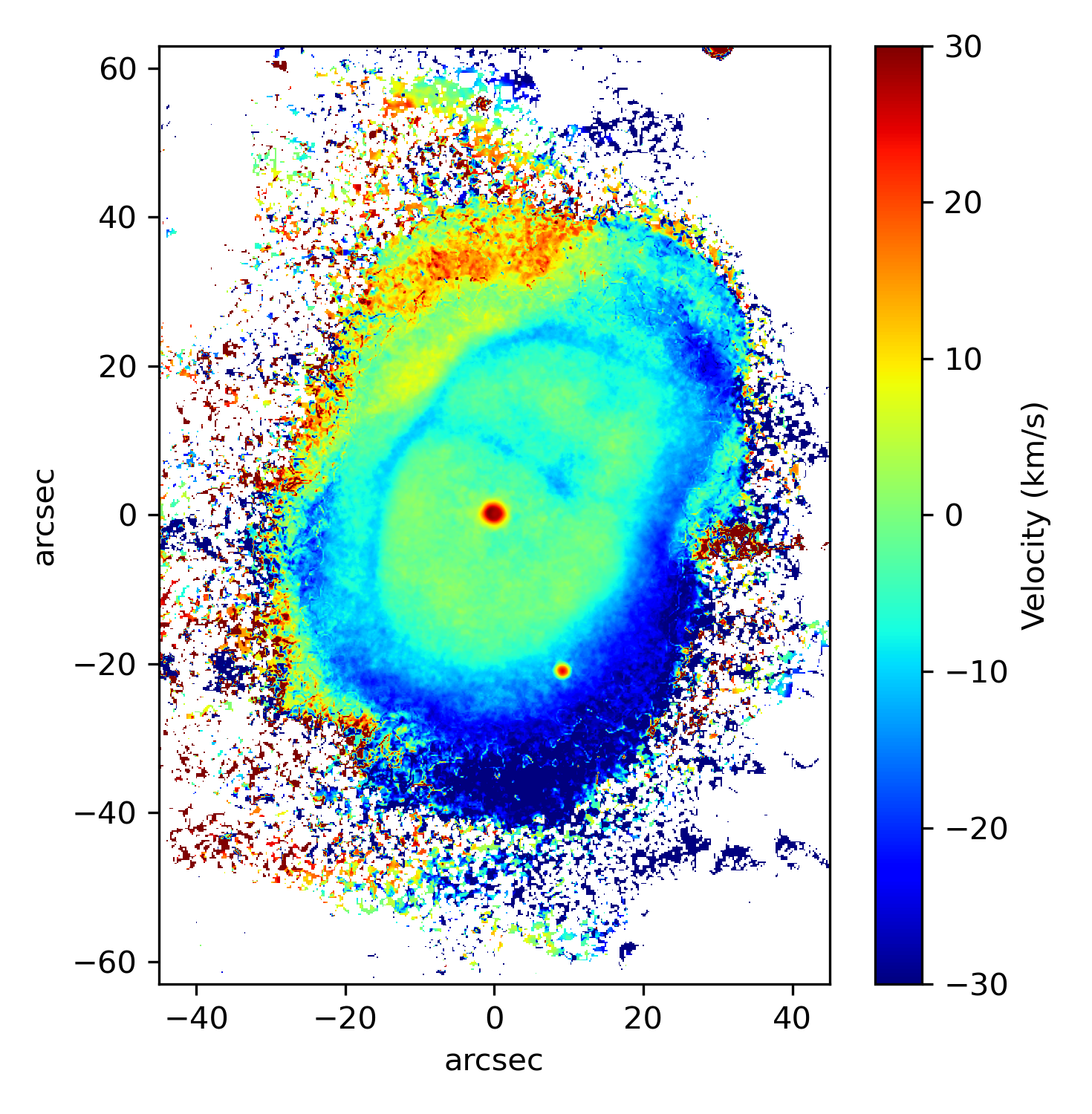}
	\includegraphics[width=\columnwidth]{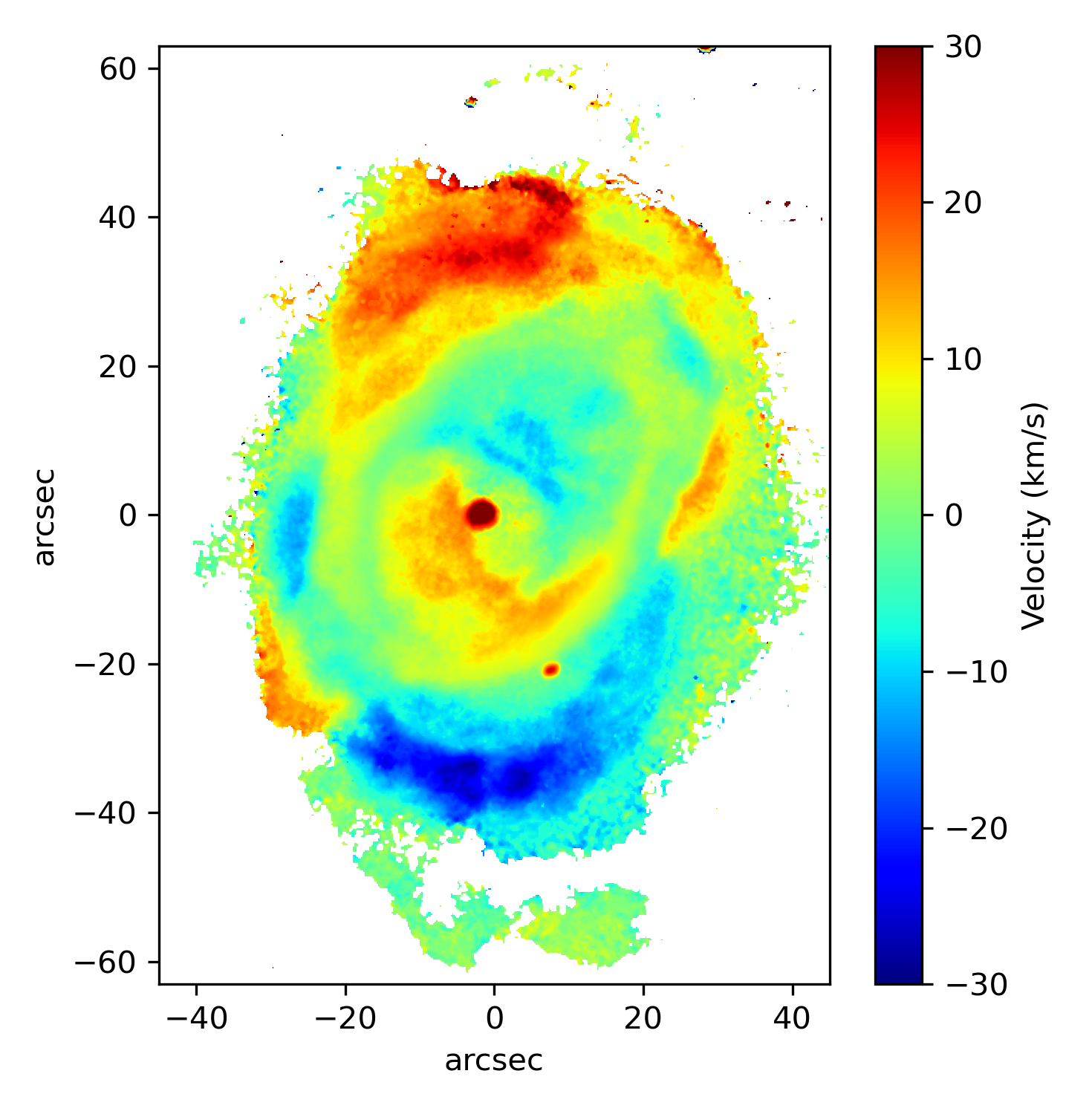}
    \caption{Velocity field maps for the velocity data cubes of 6563.85\AA~ (left) and 6584.49\AA~ (right), respectively.}
    \label{fig:velmaps}
\end{figure*}

The exposure time per channel was 30 seconds in both cases. The full width at half maximum (FWHM) of the stars, after undergoing GLAO laser correction, was estimated to be approximately 0.7 and 0.9 arcseconds, and the mean scanning lambda values for the H$\alpha$ and [N~{\sc ii}] emissions were determined as 6563.85\AA~ and 6584.49\AA~, respectively. The calibration of the observations was achieved through the utilization of a Ne~~{\sc i} reference line at 6598.95\AA~, observed with a filter with a central wavelength of 6600.5\AA~ and a width of 19.3\AA~ to isolate the Ne line. The observations were taken under photometric conditions. The FWHM of a Ne calibration lamp lines was 0.586\AA~ or 26.8~km~s$^{-1}$, which corresponds to a spectral resolution of about 11200 at H$\alpha$. The night sky lines were identified by plotting a histogram of wavelengths and picking the most frequent values, given that they are present in every pixel). Knowing their wavelengths and intensities, night sky lines were then subtracted from the cubes. 

Although the data have significantly high resolution, we opted to use deconvolution to push the limit and improve the detection of distinct velocity components. To achieve this, we used the Richardson-Lucy deconvolution algorithm implementation modules of the scikit-image restoration package\footnote{\url{https://scikit-image.org/docs/stable/api/skimage.restoration.html}}. For performing the deconvolution, we adopted a point spread function (psf) with a FWHM given by the spectral resolution of the data.

The observed velocity map is rich in structure. In general, it shows the signatures of an expanding gas with some substructures. This can be seen in Fig. \ref{fig:velmaps}, where we show maps with detailed kinematic information of the whole nebula in \ha~(left) and \nii~6584\AA~ (right), obtained by calculating the intensity-weighted velocity of the spectral line, which is the first moment of the spectral data cube. The images show a field with approaching velocities in the northern part of the nebula and receding in the southern part. In the inner regions, within what is usually referred to as the "rim" of the nebula, the \ha~map shows very little structure, whereas the \nii~6584\AA~ image shows, in the southern part, an approaching region, and in the northern part, a receding one. These results agree with those presented by \citet{MUSE3132}.

\begin{figure}
	\includegraphics[width=\columnwidth]{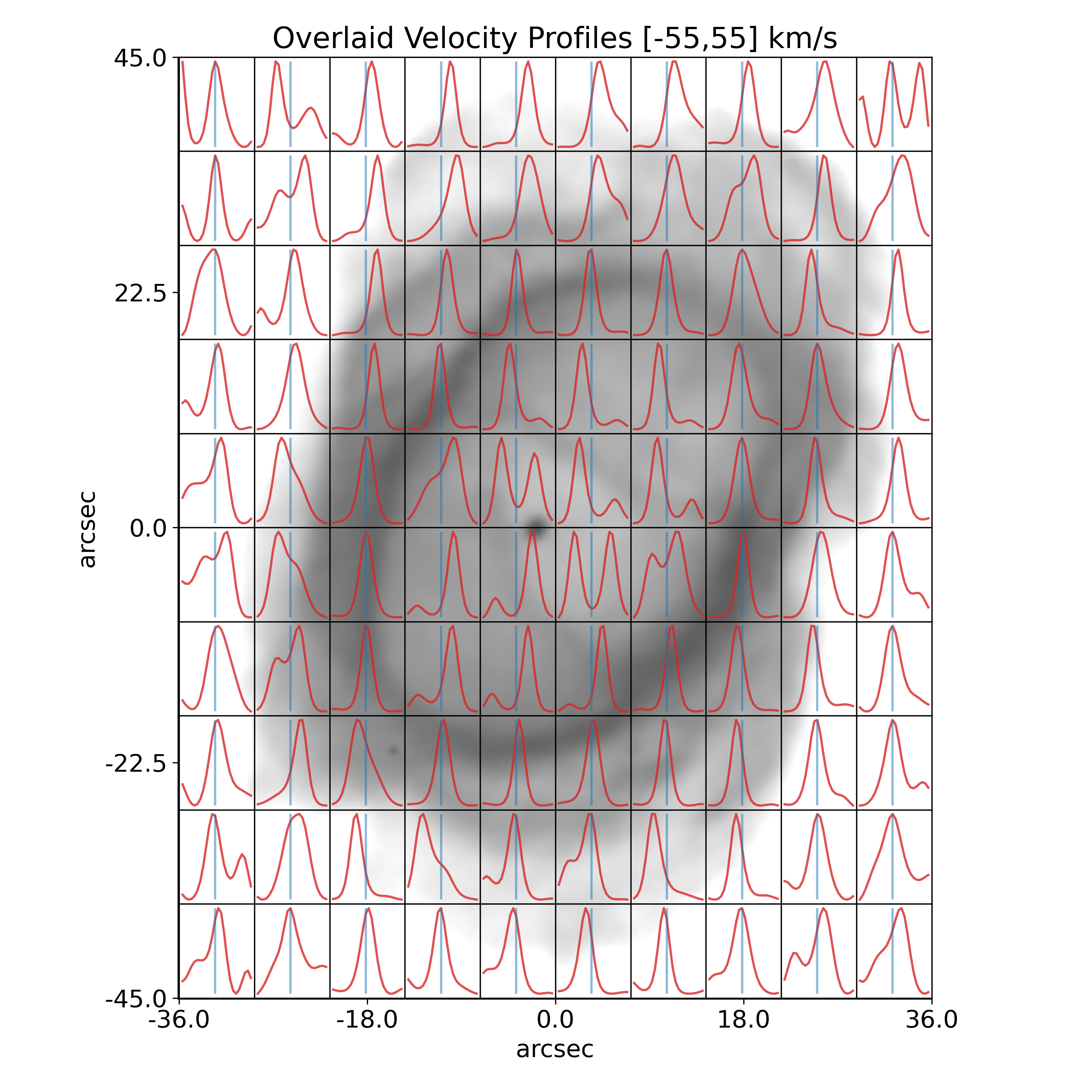}
    \caption{SAM-FP velocity profile grid for NGC 3132 obtained from the SAM-FP \nii~6584\AA~ velocity datacube, showing the velocity structure at each grid position.}
    \label{fig:velgrid}
\end{figure}

Another way to look at the velocity information in the data is to inspect the velocity profiles along a given line of sight. This is what is shown in Fig. \ref{fig:velgrid}, where we show a grid of velocity profiles overlaid on the \nii~6584\AA~ image obtained from the SAM-FP cube. Each plot of the grid shows the velocity profile obtained by summing the data for an aperture of 1", placed at the central position of the plot box. Here we see some double-peaked profiles in the very central regions of the nebula, with varying degrees of contribution from the approaching and receding parts as we get closer to the "rim" of the nebula. In the outer regions, we again see the reversal of the dominant component, as mentioned before.

\begin{figure*}
	\includegraphics[width=\columnwidth]{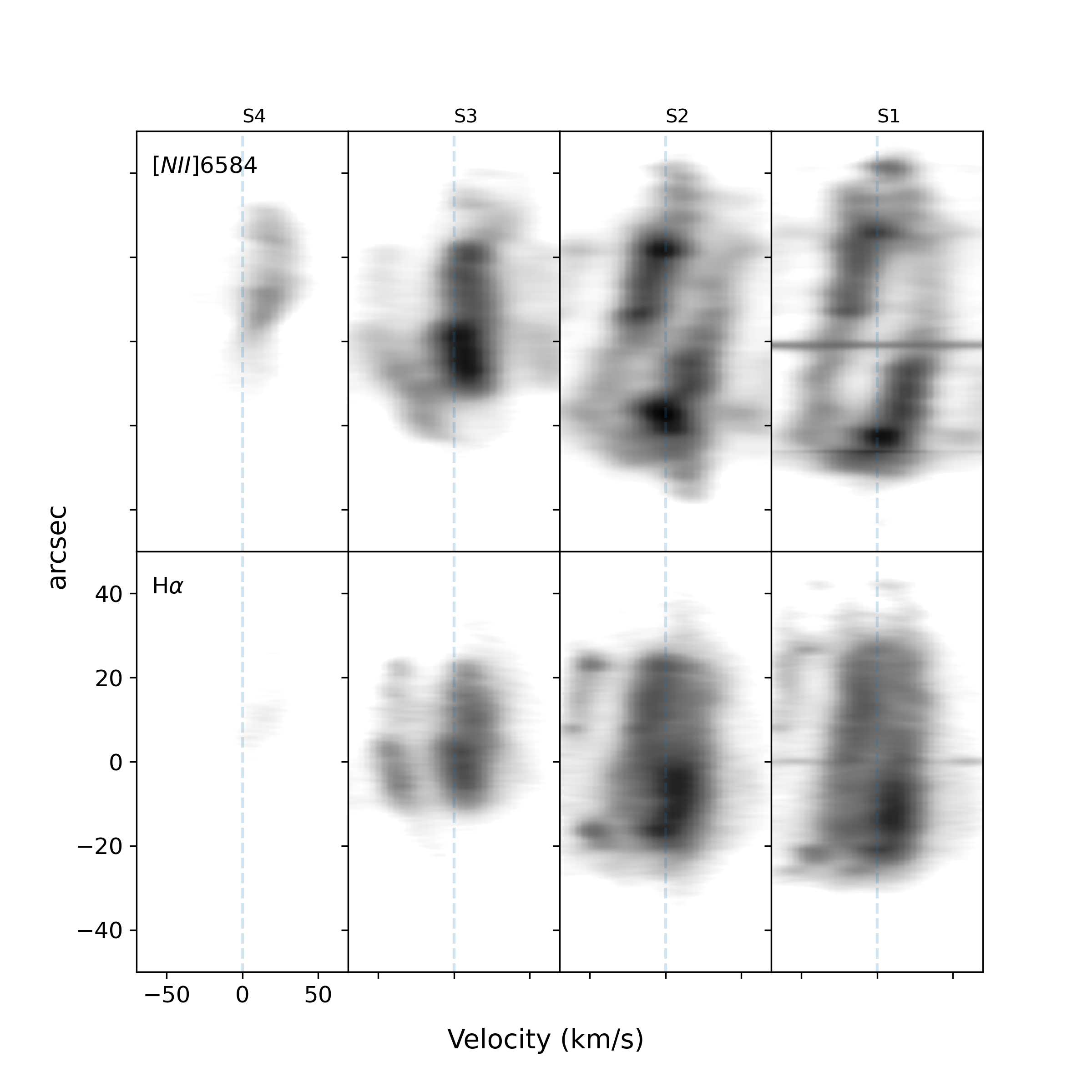} 
        \includegraphics[width=\columnwidth]{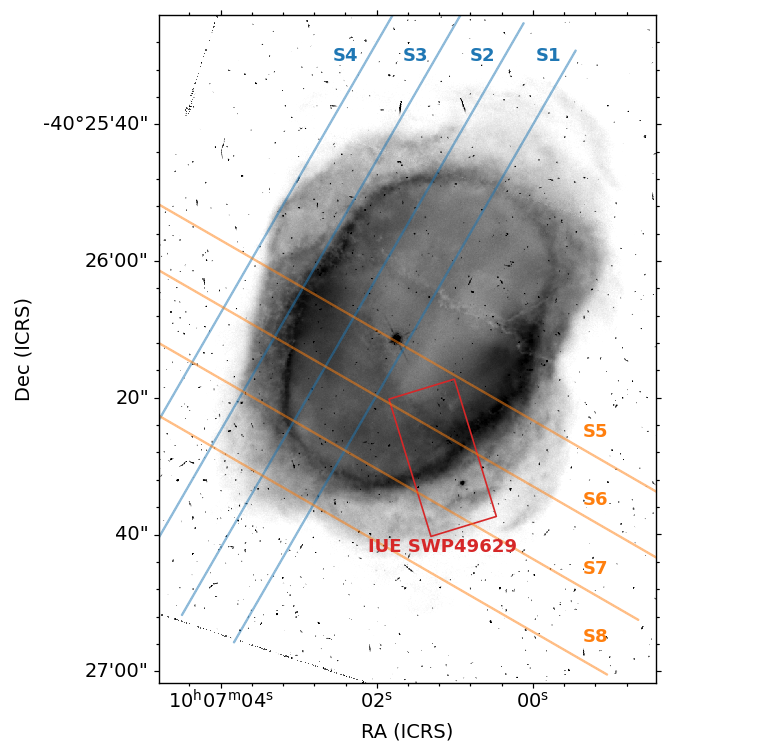} \\
        \includegraphics[width=\textwidth]{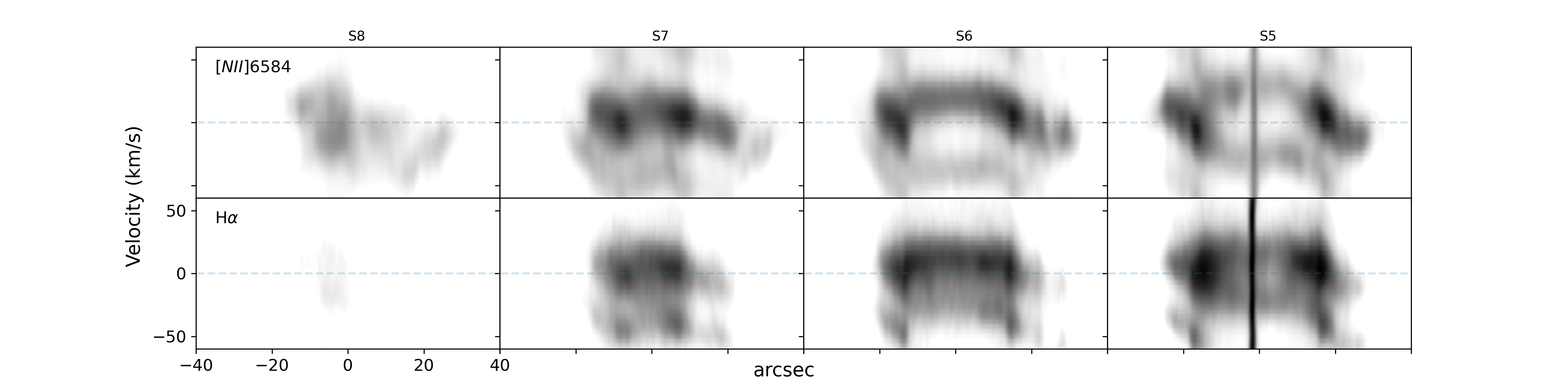}
    \caption{Position-velocity maps obtained from the SAM-FP \nii~ datacube for distinct simulated slit positions as shown in the upper right panel, where the slits are overlaid in a HST/WFPC2 \ha~ image, obtained from the Hubble Legacy Archive. The two IUE apertures used in this work, as described in Sec. \ref{sec:iuedata} are also shown.}
    \label{fig:PVdiags}
\end{figure*}

A more interesting visualization, considering the distribution of distinct velocity structures, in the case of an object like a planetary nebula, is the position velocity (PV) diagram. With the velocity data cube we simulated distinct slits to obtain PV diagrams in some regions of the nebula. The simulated slit as well as the respective PV diagram obtained for 8 distinct positions are presented in Fig. \ref{fig:PVdiags}. Four slits were positioned parallel to the major axis and four parallel to the minor axis of morphological symmetry of the nebula. We obtained PV diagrams for the \ha~ and \nii~6584\AA~ data cubes. Overall, the PV diagrams show the signature of an expanding gas bubble with some complex substructures. Looking at the PV for the S1 slit, we can see a somewhat bipolar cavity with a point-symmetric enhanced shell that does not completely surround it. The point symmetric enhancements can also be seen in the slits S2, S3, and less clearly so in the respective \ha~ PV diagrams. These structures are also seen in the data presented in \citet{Hajian2007}.

The velocity data presented, apart from being important in itself, were also used to derive the input three-dimensional gas and dust structure of the nebula in the model. This is described in detail in Sec. \ref{sec:3Dstrut}.

\subsection{Observations from IUE}
\label{sec:iuedata}

In this work, we use IUE observations in the same manner as \citet{Tsamis2003} to constrain our model carbon abundance, as they are the only observations that include collisionally excited lines for this element, which are crucial to determining its abundance. However, we also remeasured the IUE aperture spectra for  NGC~3132 that they presented. The data comprise low-resolution, large-aperture spectra obtained with the SWP camera, covering the wavelength range of 1150 to 1975 angstroms, which were accessed on the Space Telescope Science Institute website\footnote{\url{https://archive.stsci.edu/iue}}. The IUE large aperture measuring 10.3 by 23 arcseconds was used to capture the observations which were processed with the final NEWSIPS calibration procedures. In this work, we use the data from the aperture swp49629, since it was the one with the best quality available. Data for aperture swp49629 were downloaded and the emission line intensities for the detected lines were obtained with a Gaussian fit. In Fig. \ref{fig:PVdiags} we show the aperture and its positioning in relation to the nebula.

To be able to use the line-intensity data of these observations, combined with intensities measured with other instruments in the optical to constrain the model, we need to scale the UV calibration to the optical one, given that observing procedures in each are very distinct. To achieve this, we used the emission line \heii~$\lambda$1640 from the UV and \heii~$\lambda$5411 from the optical MUSE data and the fact that the ratio of these lines is theoretically well determined.

Unlike \citet{Tsamis2003}, we scaled the observations without making assumptions about aperture corrections, using the spatially resolved nature of the MUSE data to extract the flux of the \heii~$\lambda$5411 in the same aperture configuration as that used for the IUE observations. This procedure guarantees that we are not introducing incorrect scaling factors due to the distinct ionisation structure of the lines involved and positioning of the aperture. To determine the theoretical ratio, we use our model to calculate it under the exact conditions of the nebula and with the same aperture configuration as in the observations.

The final values we obtained and used as constraints on the model, were \heii~$\lambda$1640=(0.851$\pm$0.113)\hb~and \textsc{C\,iii]}$\lambda\lambda$1906+1908=(2.642$\pm$0.135)\hb. The values determined by \citet{Tsamis2003} were \heii~$\lambda$1640=0.259~\hb~and \textsc{C\,iii]}$\lambda\lambda$1906+1908=0.411~\hb.





\section{3D photoionization model}
\label{sec:3Dmodel}
To investigate the ionization structure and construct detailed models of PNe with complex morphologies such as NGC 3132, sophisticated 3D photoionization codes are necessary. In this work, we use the MOCASSIN photoionization code (version 2.02.73.2), as detailed in \citep{2003MNRAS.340.1136E, 2005MNRAS.362.1038E}. The code utilizes atomic data from the CHIANTI database (version 10), as described in \citet{2021ApJ...909...38D}. The overall process of modelling a PN with MOCASSIN involves creating a density distribution of the nebular gas and dust and running a set of models varying the input central source luminosity and effective temperature as well as the elemental abundances until a satisfactory fit to the predefined constraints is achieved. A similar three-dimensional modelling approach with MOCASSIN has also been used for the study of the planetary nebulae Abell~14 \citep{Akras2016}.  In the following, we describe in detail the methodology and assumptions adopted in each of those steps.

\subsection{The gas density structure}
\label{sec:3Dstrut}
One of the key elements in constructing a detailed photoionisation model for a PN is the definition of the input density structure (gas or dust). The most common procedure is to assume a simple one-dimensional structure with spherical symmetry. More detailed options such as hydrodynamical models, while providing self-consistent results, rely on numerous assumptions about the underlying physical processes without direct observational constraints beyond observed morphologies. Structures obtained in this way are good for studying generic PNe types, such as bipolars, for example; however, they are not ideal for specific objects unless a dedicated model is constructed. In contrast, kinematical modelling approaches \citep[e.g.][]{Akras2016}, which use a limited number of high-resolution slit observations, often depend on idealised assumptions about the 3D structural elements being combined, such as cylinders, toruses, cones, and other geometric shapes. These idealisations may not accurately represent the complex and varied morphologies of actual PNe. 

In this work, we make use of the spatially resolved velocity information contained in the SAM-FP data cubes discussed in Sec. \ref{sec:samfp} to infer a three-dimensional structure for NGC 3132. The procedure is based on the assumption that the velocity of a given volume element of the gas is constant over the expansion time of the nebula. According to \citet{Zijlstra01}, this type of velocity distribution can be expected if the nebula has evolved from a relatively short mass loss event and is now moving ballistically. In addition, \citet{Steffen04} argue that a continuous interaction of a wind with small-scale structures can also develop a positive velocity gradient with distance. The result is a homologous expansion in which the nebula conserves its shape over time.

In practical terms, the velocity vector at any given point is proportional to the position vector of that point in the nebula. Evidence of this property can be found in the literature \citep[among others]{Wilson1950,Weedman1968, 2004ASPC..313..148C, 2008MNRAS.385..269M, 2014MNRAS.442.3162U}. This assumption has been used in numerous works, where the goal was to deproject the spatial structure of PNe. Perhaps the most focused on the technique is the work of \citet{Sabbadin06} and references therein, where the group obtained the three-dimensional structures of a sample of objects using high-resolution long-slit spectroscopy. In the work of \citet{Shape06} the authors present a computational tool to analyse and disentangle the three-dimensional geometry and kinematic structure of gaseous nebulae called SHAPE\footnote{\url{https://wsteffen75.wixsite.com/website}}, which is still widely used in the literature to study PNe, recovering their detailed kinematical and structural information \cite[e.g.][]{Vayet2009,Clark2010, akras2012a,Akras2012b,Clark2013,Akras2015,Clyne2015,Harvey2016,Derlopa2019,Derlopa2024}

To derive the density structure for NGC 3132 based on the SAM-FP data and the assumptions discussed previously, we began with an initial estimate of the proportionality constant used to convert velocity field information into distances. We also assumed that, as a first approximation, the density of the gas is related to the H$\alpha$ emission intensity as I$_{\text{H}\alpha}$~$\propto$~$n_H^2$ throughout the nebula, where $n_H$ is the hydrogen gas density. By applying the velocity-to-position proportionality, we transformed the H$\alpha$ velocity data cubes into a 3D density structure. With this structure, we began the iterative fitting process of the photoionisation model to match the available observational constraints. After achieving a preliminary fit, we extended the analysis by relating the [N~{\sc ii}]$\lambda$6584 line intensity to the density. Here we also make use of the assumption that $I_{[\text{N\sc ii}]\lambda6584}~\propto$ $n_H^2$ in the regions where the line is formed, i.e. where the ion N$^+$/H is present, given that the densities are not close to the critical density of $log(N_c)=4.9$ for the line.  This relation allowed us to convert the [N~{\sc ii}]$\lambda$6584 velocity data cubes into a corresponding 3D density structure, which was then combined with the previous H$\alpha$ structure.

\begin{figure*}
	\includegraphics[scale=0.25]{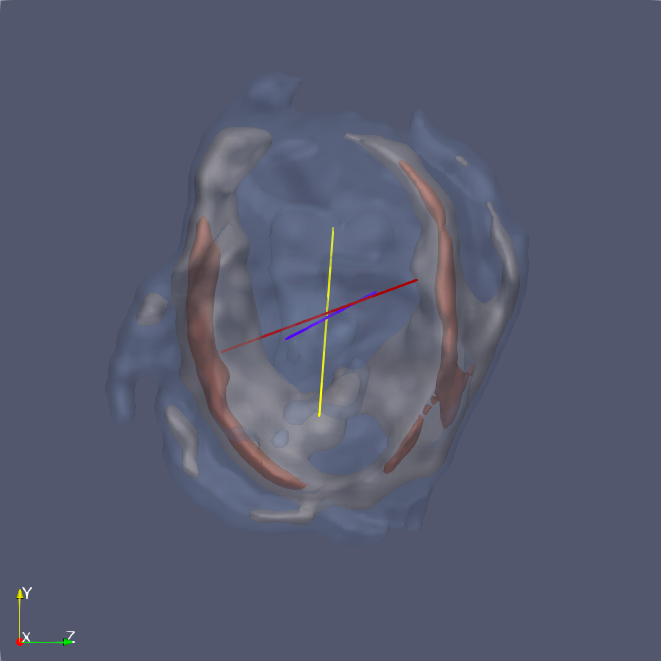} 
	\includegraphics[scale=0.25]{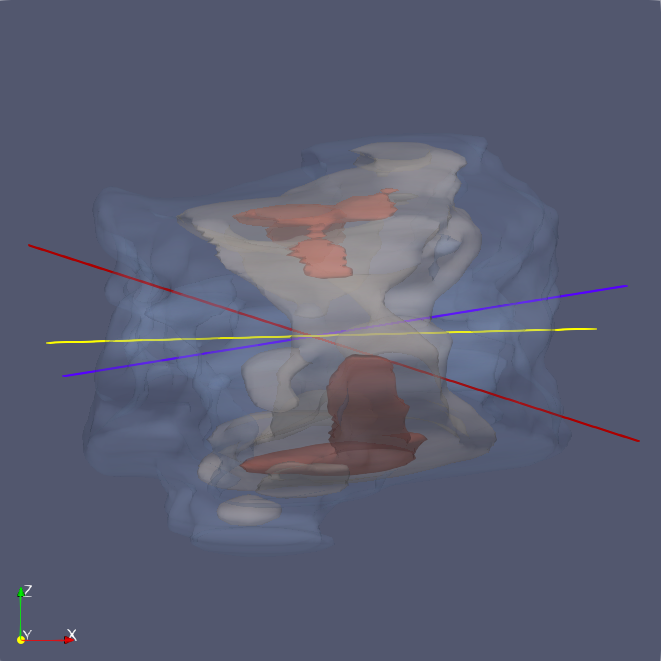} 
	\includegraphics[scale=0.25]{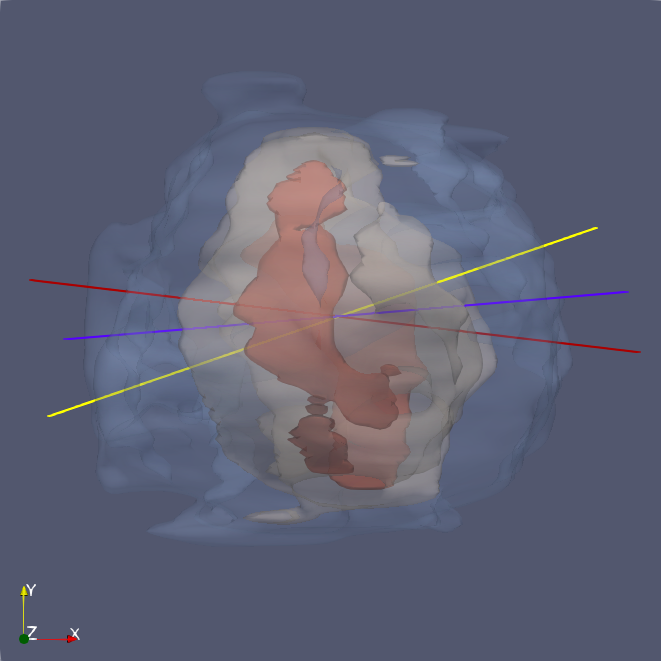} \\
 
	\includegraphics[scale=0.25]{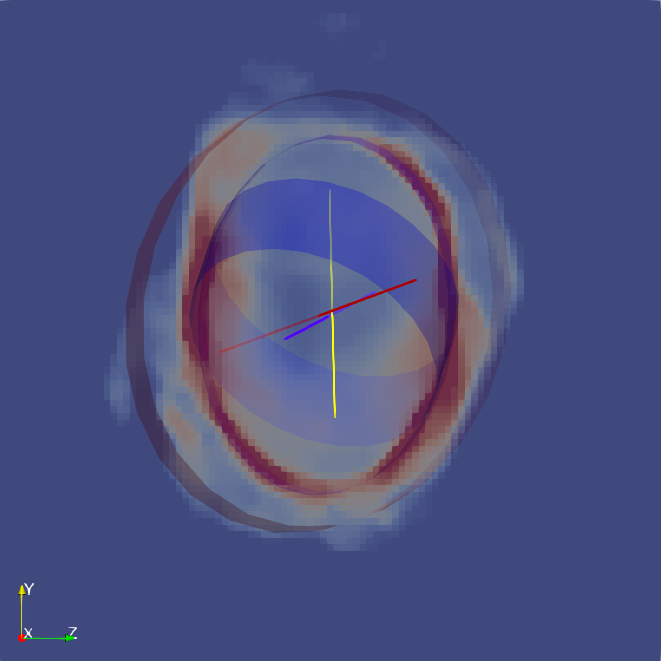} 
	\includegraphics[scale=0.25]{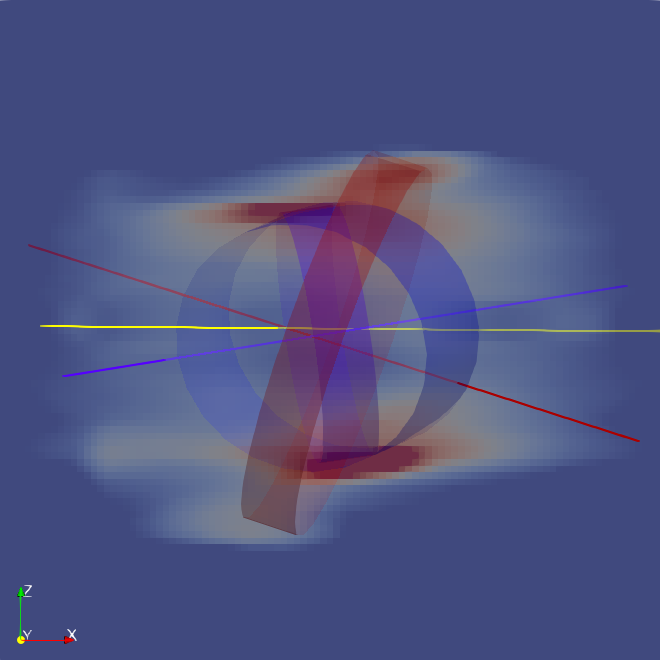} 
	\includegraphics[scale=0.25]{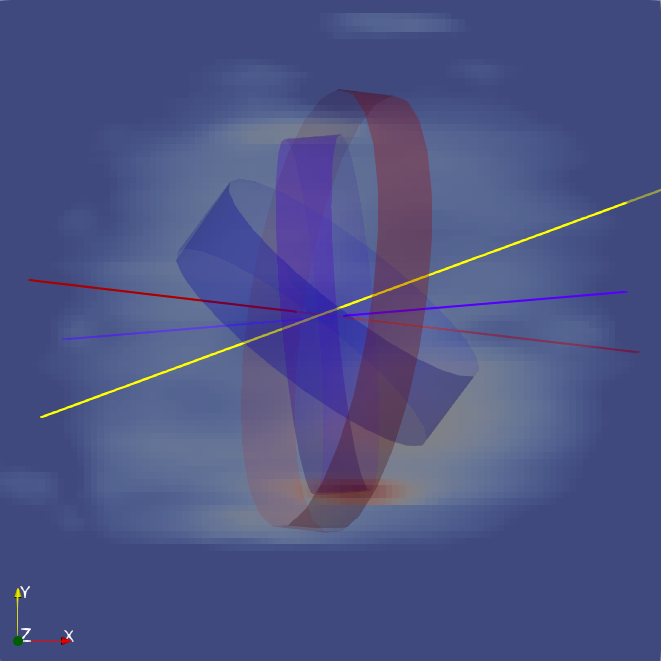} \\
 
    \caption{A visualization of the 3D structure showing in the upper row isodensity surfaces for 500 (blue), 850 (gray) and 1400 (red) $cm^{-3}$ of the density distribution obtained for NGC 3132 along the x, y and z axes respectively. Locations of the three main ring regions found are shown in the lower row along with slices of the density structure in the yz, xz and xy planes respectively. Also shown are the identified symmetry axis for the large outer ring (red line), dense inner ring (purple line) and an inner low density cavity (yellow line). }
    \label{fig:3D-dens}
\end{figure*}

To construct the combined three-dimensional density structure of the nebular gas distribution, we make use of the \nii~ and \ha~ velocity data cubes, to define the low ionisation region of the nebula, by taking the ratio $\hat{I}_{[\text{N\sc ii}]}/\hat{I}_{\text{H}\alpha}$, where $\hat{I}$ represents the intensity normalised so that max($I$)=1 for the relevant line. The resulting array is then used as a weighting factor, to differentiate regions within the nebula based on their ionisation properties. Finally, the three-dimensional density structure of the nebula was determined by combining the density arrays determined from the \nii~ and \ha~ data cubes into a single density array, weighted by the $\hat{I}_{[\text{N\sc ii}]}/\hat{I}_{\text{H}\alpha}$~array. The weight makes it so that in the inner regions, where hydrogen is completely ionised, the \ha~structure is more prevalent, while in the low ionisation zones the \nii~structure is predominant. 

In the process of finding the best model, we identified the need to introduce a filling factor, which indicates how much of the nebula's volume is occupied by gas,  of $\epsilon=0.65$ to adequately reproduce the main bright nebula, as well as the fainter outer regions. The filling factor was applied by randomly setting cells in the final density grid,  so that 35\% of them were set to zero density.

We also had to increase the density in low ionisation zones in particular where N$^+$ starts recombining and the dependence $I_{[\text{N\sc ii}]\lambda 6584} \propto n^2$ starts to no longer be valid (see Fig. \ref{fig:ionic_profiles} for details. With the help of the model, we determined that the zone where the temperature dependence starts to become important for the collisional lines corresponds to $\hat{I}_{[\text{N\sc ii}]}/\hat{I}_{\text{H}\alpha}>0.7$. For regions above this threshold, we increased the density according to the relation $(\hat{I}_{[\text{N\sc ii}]}/\hat{I}_{\text{H}\alpha}-0.7)\times700~cm^{-3}$, yielding an increase of about 1400$~cm^{-3}$ in the very low ionisation zones where $\hat{I}_{[\text{N\sc ii}]}/\hat{I}_{\text{H}\alpha} \approx 2$. The final density distribution was obtained by:

\begin{align}
    R = \frac{\hat{I}_{[\text{N\sc ii}]}}{\hat{I}_{\text{H}\alpha}} \\
    N_\text{H} = 1350\sqrt{\hat{I}_{\text{H}\alpha}}\times(1-R) + 1800\sqrt{\hat{I}_{[\text{N\sc ii}]}} \times R + C \\
    C = 
\begin{cases}
    (R-0.7)\times700~cm^{-3} & \text{if } R > 0.7 \\
    0 & \text{if } R \le 0.7
\end{cases}
\end{align}

\begin{figure*}
	\includegraphics[width=\textwidth]{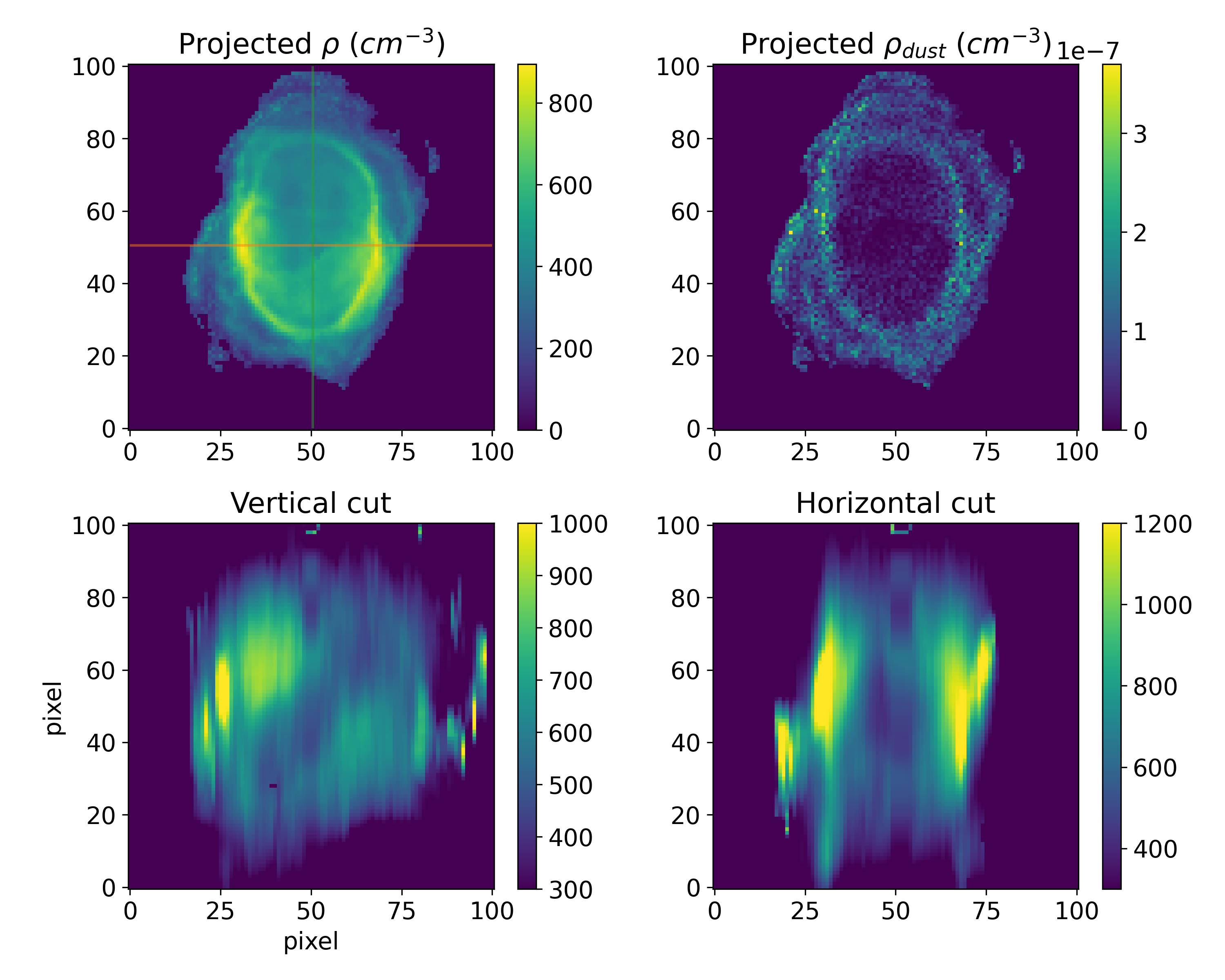}
    \caption{Visualizations of the dust and gas density structure adopted. Shown are projected gas density (upper left), projected dust density (upper right), major axis cut of the gas density (lower left) and minor axis cut (lower right).  }
    \label{fig:strutcuts}
\end{figure*}

A visualisation of the resulting 3D density structure is provided in Fig. \ref{fig:3D-dens}. In the figure, we show isodensity surfaces for 500 (blue), 850 (gray) and 1400 (red) g~cm$^{-3}$ of the density distribution obtained for NGC 3132 along the x, y, and z axes, respectively. Inspecting this density structure in the 3D visualisation we can identify some prominent structures: a mid-density outer ring, two higher density inner rings, and a low-density region like a cavity. The main dense elliptical inner ring has a symmetry axis (purple line in Fig. \ref{fig:3D-dens}) which is inclined relative to the line of sight by about 30 degrees. This denser inner ring is not a regular structure, and some warping can be seen. The second denser inner ring is of slightly lower density and is also not contiguous, showing a less defined structure (perhaps also due to data quality). This secondary inner ring is also related to some of the more obvious areas of higher extinction (see, for example, Fig. \ref{fig:PVdiags}). The outer lower density ring shows a more irregular structure, with breaks and warps perhaps due to interactions with previously ejected material as well as the interstellar medium. Its shape is also elliptical but with less eccentricity. The outer ring is tilted in relation to the denser inner ring (lower row in Fig. \ref{fig:3D-dens}) and the transition in densities between the two is well defined but continuous, as can be seen in Fig. \ref{fig:strutcuts}. The figure also shows identified symmetry axis for a large outer ring (red line), dense inner ring (purple line) and an inner low density cavity (yellow line).

\begin{figure}
    \centering
    \includegraphics[width=\columnwidth]{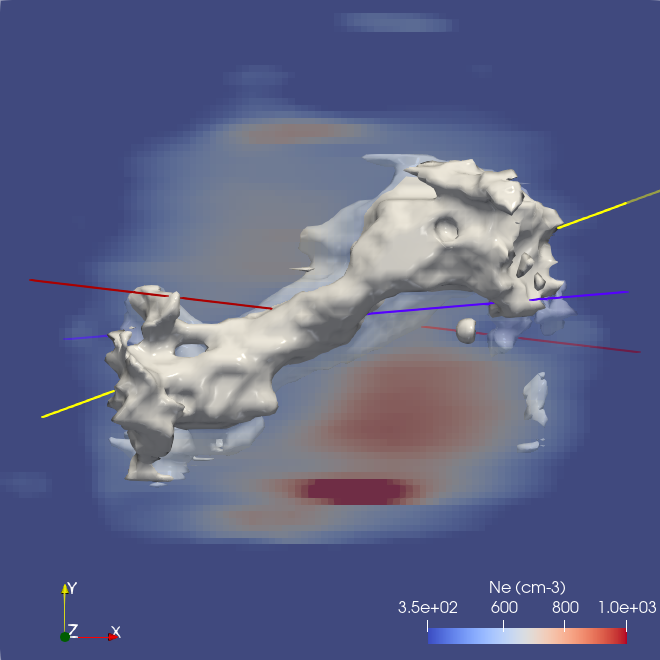}
    \caption{Isodensity contour showing the inner low density cavity together with a slice in the yx plane of the gas structure. }
    \label{fig:cavity}
\end{figure}

The low density cavity (400 - 550 g~cm$^{-3}$) shows a complex somewhat warped bipolar shape when looking at a cut through the density structure along the major axis of the projected morphology (Fig. \ref{fig:cavity}). The cavity runs through the nebula along an axis inclined by about 20 degrees in relation to the line of sight.

\subsection{The dust density structure}

For the NGC 3132 model, we also included dust in the calculations. This is justified by infrared observations of the outer regions of the nebula as well as in the inner regions, as evidenced by recent JWST images presented in \citet{2022NatAs...6.1421D}, and also due to the importance of dust in the energy balance of the system. Although dust was not the main focus in this work, we experimented with a wide range of dust distributions in relation to the gas structure, from uniform to locally distributed. In the fitting process, we found that the distribution that gave the best results, with respect to how well the model reproduced the observed spectral energy distribution (SED) in the infrared (see Fig. \ref{fig:SED}, was non-uniform and with dust present in the low ionisation zones of the nebula. 

The final configuration of the dust structure that gave the best results was generated by adopting a grain number density proportional to the H number density of the gas,  present only in regions where $\hat{I}_{[\text{N\sc ii}]}/\hat{I}_{\text{H}\alpha}>0.25$, in other words, with a constant dust-to-gas ratio by number in the region where dust is present. The final model dust distribution in the main nebula has a total mass of $1.0\times10^{-2}M_{\odot}$, giving an overall dust-to-gas mass ratio of 0.07.

\subsection{The central ionising source}

The ionising source, characterised by its $T_{eff}$ and luminosity, is one of the essential free parameters that we fit in the model. Initially, we used standard blackbody sources, but these failed to fully replicate the observations, in particular the gas temperature measured through diagnostic line ratios, always producing significantly hotter plasmas. Accurately reproducing the gas temperature is key in ensuring that the abundances derived from the model are reliable. Consequently, we experimented with more sophisticated NLTE stellar model atmospheres from T. Rauch\footnote{\url{http://astro.uni-tuebingen.de/~rauch/}}, which yielded improved results but still produced gas temperatures a few hundred degrees hotter than those indicated by observational diagnostics.

The results of the JWST telescope observations presented in \citet{2022NatAs...6.1421D}, where a dusty envelope surrounding the central star of the nebula was detected, provided important observational constraints, including precise photometry of the central source. In that work, they reported observations consistent with a dust disk having an inner radius of 55 AU, an outer radius of 140 AU, and a dust mass of $2 \times 10^{-7}\textrm{M}_{\odot}$. Similarly, in \citet{2023ApJ...943..110S}, using the same dataset, the authors concluded that a dust mass of $3.9 \times 10^{-8}\textrm{M}_{\odot}$, extending to a radius of 1785 AU and composed of 70\% silicate and 30\% amorphous carbon, could match the observed photometry. However, as discussed in \citet{2022NatAs...6.1421D}, a dust-only shell or disk alone was insufficient to improve the temperature predictions of the photoionisation model. 

These results led us to consider a combined dust and gas shell surrounding the central source in our models. To achieve this, we introduced a spherical shell of a uniform density of dust and gas around the ionising source. In principle, this region could have been handled by the code using the nested grid capacity of Mocassin; however, due to the large number of packets required for the convergence of the central region alone, the computational cost of doing this self-consistently with the main nebula would become prohibitive with our available computational resources. We chose then to separate the problem into main nebula and central region and run models separately for both. This allowed us to run models more quickly and converge faster to the solution for the nebula and for the central region without the added computational time and complexity of large nested grids with multi-chemistry for dust and gas. With this we first generated a photoionisation model for the central region, then used the resulting spectral energy distribution (SED) as the input ionising source in the photoionisation model of the main nebula.

By iterating over the shell's free parameters such as size, inner and outer radius, density, and composition, we found a model configuration that better reproduced line diagnostic ratios, particularly for temperature. The best fit was obtained with a spherical shell with inner and outer radius of 67 and 334 AU respectively, containing $2.42 \times 10^{-5}\textrm{M}_{\odot}$ of gas and $2.42 \times 10^{-10}\textrm{M}_{\odot}$ of dust that gives a dust to gas ratio by mass of $1 \times 10^{-5}$. We also note that the fit was sensitive to the shell's elemental abundances. The best match was obtained with a He-poor, C- and O-rich composition, with mass fractions of $X_\text{He} = 6.5\times10^{-3}$, $X_\text{C} = 0.474$, $X_\text{N} = 4.7\times10^{-7}$, and $X_\text{O} = 0.466$. To reach these values, we explored different combinations consistent with predictions from stellar evolution models for the inner layers of a 3M$_{\odot}$ progenitor star remnant \citep[e.g.,][]{2016ApJS..225...24P}. 

The dust composition was also constrained due to the central source photometry available from \citet{2022NatAs...6.1421D} and \citet{2023ApJ...943..110S}. We found that the photometry was best reproduced by dust consisting of graphite grains with radii $0.2\micron$ to $1.0\micron$ \citep[data for grains from][]{1984ApJ...285...89D}.

The final central source spectrum is shown and discussed in detail in sec. \ref{sec:results}.

All files used in generating the final Mocassin model have been made available at \url{https://github.com/hektor-monteiro/NGC3132_model}. There the reader will also find a Python notebook which can be use to visualize isodensity surfaces and slices of the density structure.


\subsection{Finding the best model}
\label{sec:optimize}

{During the modelling process, we used a two-stage approach to refine the photoionisation model of NGC 3132. In the first stage, for which we gave details in previous sections, we follow a traditional method by manually running models and comparing them against all available observables, including emission-line maps and integrated line fluxes and diagnostic ratios. This step ensures that key structural parameters, such as the \hb~flux, nebular size, and energy balance of the nebula, are properly constrained. It also ensures that the stellar effective temperature and luminosity, which directly influence the overall physical conditions of the nebula, are already well within accepted tolerances. The density distribution is constrained by the available emission line intensities and diagnostic ratios, in particular the spatially resolved images, to ensure that the spatial properties are well reproduced. Since density directly affects the obtained model \hb~flux, this value is also constrained in this stage by the observed value for the entire nebula.

In principle, some degeneracy is expected between density, size, and central source luminosity, especially when the distance adopted is uncertain and the model considers a filling factor. In our case, since the distance, and therefore the size of the nebula, is really well constrained by the distance value obtained from GAIA, that problem is minimised. Furthermore, spatially resolved data for emission lines from distinct elements and ionisation stages put strong constraints on the luminosity of the central source and the resulting model \hb~flux at this stage. The results from the final model and how well it reproduces these constraints are discussed in detail in Sec. \ref{sec:results}.

With the density of gas and dust as well as the properties of the central ionising source defined and constrained by the observational data as described previously, we then focus on the second stage which is the fitting of the elemental abundances of the key elements He, C, N, O, and S. Although all abundances and properties mentioned before are usually the free parameters of a model, which are determined by the fitting process constrained by the available observations, in this second stage, we adopt the central ionising source and density structure parameters as fixed since they already are constrained within the adopted tolerances.    

In this second stage, to determine the best fit of the key elemental abundances of the model, we use a global optimisation algorithm called cross-entropy, which has been successfully applied to a series of astrophysical problems, as in \citet{2017ApJ...851L..39C} and \citet{2021MNRAS.504..356D} and references therein. The main numerical constraints adopted at this stage for the fitting of the abundances are the integrated emission line intensities and the derived diagnostic ratios. The Cross Entropy method involves an iterative process in which the following is performed in each iteration:

\begin{enumerate}
\item random generation of an initial sample of fit parameters,
\item selection of the $10\%$ best candidates based on calculated likelihood values,
\item generation of a random fit parameter sample, derived from a new distribution, based on the $10\%$ best candidates calculated in the previous step,
\item repeat until convergence or stopping criteria reached.
\end{enumerate}

We have written a series of Python scripts implementing the use of the cross-entropy, with the photoionisation code Mocassin, to perform model fits. A key step at this stage is the definition of the likelihood function to be adopted for the problem. Given that the main constraints for the abundance determination are the integrated emission line intensities and their respective diagnostic ratios and that their uncertainties are essentially Gaussian distributed, we adopt the following likelihood function: 

In our code, we adopt a log-likelihood function given in the usual manner, for the
maximum likelihood problem, as:
\begin{equation}
\ln \mathcal L (D|{\bf X}) \propto  - \sum_{i=1}^{n} \frac{(I_i - M({\bf X}))^2}{2 \sigma_i^2}.
\label{eq:likelihood}
\end{equation}

where ${\bf X}$ is the vector of parameters (in our case the elemental abundances to fit) of the model $M({\bf X})$, $I_i$ are the observed line intensities and diagnostic ratios to fit with their respective $\sigma_i$ uncertainties. The optimisation algorithm then finds ${\bf X}$, which maximises $\mathcal L (D|{\bf X})$.

In this optimisation stage, we kept the free parameters limited to the abundances of He, C, N, O and S, since these were the elements with the highest impacts in the temperature balance of the nebula and for which we had good observational constraints. Abundances for the other elements, such as Ar, Cl, Ne, among others, were just adjusted afterwards to obtain an adequate line intensity match to the available observations. The parameter space for the optimisation was as follows:

\begin{itemize}
    \item He: $0.09 \leq \mathrm{He} \leq 0.13 $
    \item C: $8 \leq \mathrm{C} [10^{-4}] \leq 12 $
    \item N: $5 \leq \mathrm{N} [10^{-4}] \leq 5 $
    \item O: $2 \leq \mathrm{O} [10^{-4}]\leq 8 $
    \item S: $0.2 \leq \mathrm{S} [10^{-5}] \leq 2$
\end{itemize}

The highest resolution model we can run in our current facility consists of a grid with $101^3$ cells, and models in this resolution take many hours to converge. To make the use of computation time more efficiently, we optimize the problem using model grids with $71^3$ resolution. This configuration allows for an optimization to be complete in 6-8 days of computing. The final model is then run at full resolution with the best-fit elemental abundances.

\section{Results}
\label{sec:results}

In the previous section, we described the general methodology adopted for constructing the photoionisation model for NGC 3132. Some results related to the density structure were already presented there, and here we focus on other results that can be obtained from the best-fit model, in particular relating to the chemical composition.

One of the main goals of the modelling efforts is to obtain estimates of chemical abundances. In Fig. \ref{fig:likelihoods} we show the results obtained by the optimisation detailed previously to find the best solution for the chemical abundances He, C, N, O and S, with the abundances of C, N and O normalised by $1\times10^{-4}$ and S by $1\times10^{-5}$ for clarity. The histograms show marginalized distributions, with the mean values for each abundance as well as the 1$\sigma$ confidence intervals. The 2D distributions show the calculated model points in blue, the best solution in orange and the 1, 2 and 3$\sigma$ confidence intervals. The likelihood space is regular, with no obvious correlations present. The best solution found by the algorithm, shown as an orange point, and the mode of the distributions agree well.

\begin{figure*}
    \includegraphics[width=\textwidth]{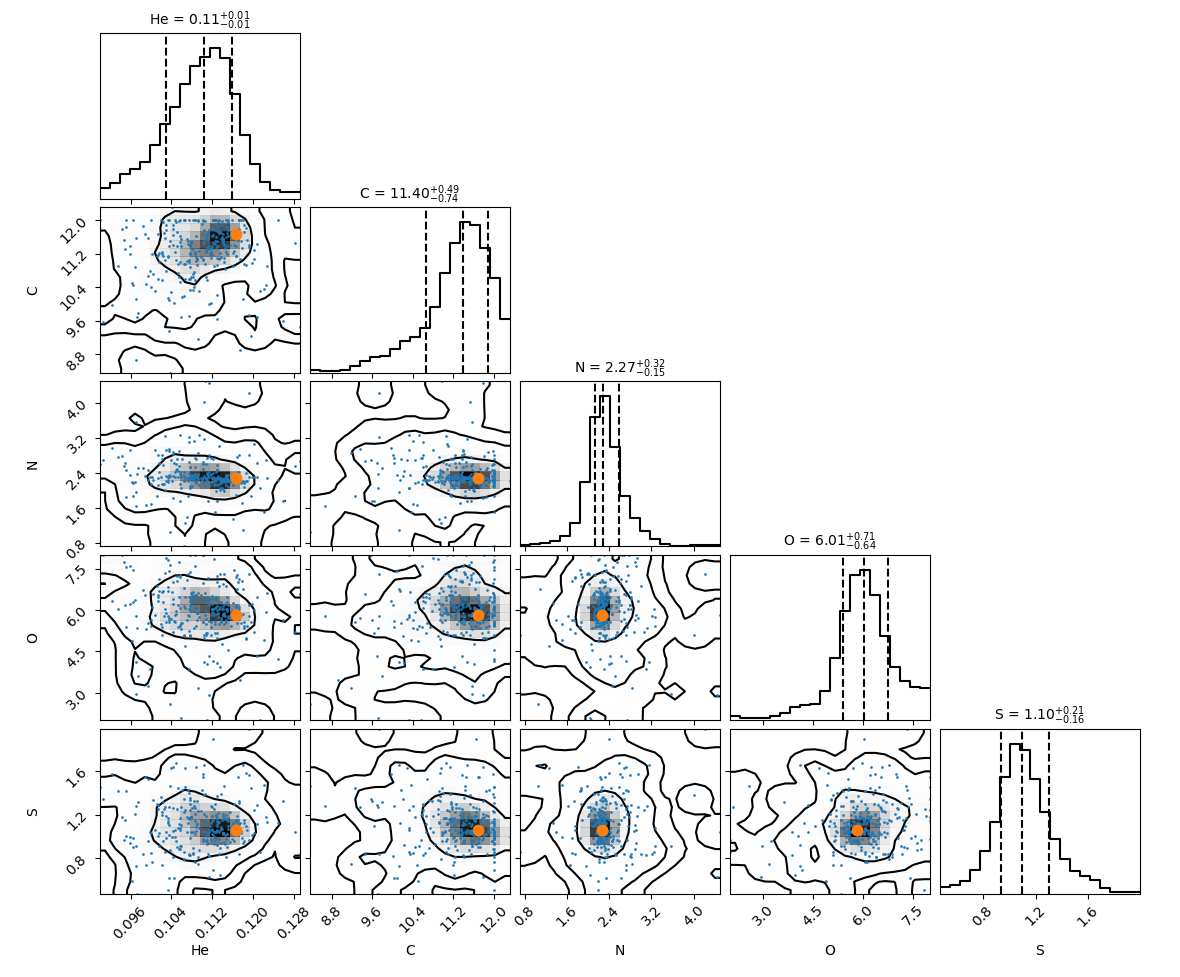}
    \caption{Results of the optimization showing the best solution found and their respective confidence intervals for the abundances of He, C, N, O and S. The C, N and O abundances were normalized by $1\times10^{-4}$ and S by $1\times10^{-5}$ for clarity. The histograms show marginalized distributions with the mean values for each abundance as well as the 1$\sigma$ confidence intervals. The 2D distributions show the calculated model points in blue, the best solution in orange and the 1, 2 and 3$\sigma$ confidence intervals.}
    \label{fig:likelihoods}
\end{figure*}

Establishing the goodness-of-fit of a given model is important, and \citet{2023arXiv231201873S} suggests that each of the mismatched line intensity has a relevant significance and suggests that the use of a quality factor $\kappa(O)$ is more appropriate. The quality factor, defined in \citet{2009A&A...507.1517M}, represents the accepted tolerance, which accounts for observational uncertainties in flux ratios and reddening, as well as the expected accuracy of the model in replicating the observations. The factor $\kappa(O)$ is defined as: 

\begin{equation}
\kappa(O) = \frac{\log(O_\mathrm{mod}) - \log(O_\mathrm{obs})}{\tau(O)},
\label{eq:QF}
\end{equation}

\noindent  where, $O_\mathrm{mod}$ and $O_\mathrm{obs}$ are the observed and modelled values of the observable and $\tau(O)$ the tolerance factor for the observable. The tolerance factor is given by $\tau(O) = log\left(1 + {\Delta O}/{O}\right)$ for any quantity $O$ with adopted tolerance of $\Delta O$.
 
For our model fits, we adopted the tolerance as the maximum of three uncertainty estimated values: 1) 3$\sigma$ from literature flux uncertainty; 2) adopted a flux uncertainty assuming that the overall calibration uncertainties are in the range 5\% to 10\%, given that absolute calibrations were not performed; and 3) adopted a 30\% flux uncertainty for lines known to be affected by other effects, such as telluric contamination or aperture effects (among others). For the infrared lines used, we adopted the value of 30\% for the flux uncertainty based on the results of \citet{2000ESASP.456..267G}.

The integrated line intensities are the main constraints used in the model fitting process in all stages and in Table \ref{tab:linecomps} we present the observations and the usual line diagnostic ratios used as constraints, their uncertainties, the relative tolerances adopted, the values obtained by the best model and the quality factor $\kappa(O)$ obtained. In general, most lines are within the adopted tolerance, indicating the quality of the fitted model. The observed absolute \hb~flux in particular, obtained by the MUSE observations, which we adopt as a constraint for this quantity, is also well reproduced. However, there are important discrepancies that are discussed below.

Three of the largest discrepancies appear for the infrared lines \oi~63$\mu m$, \oi~146$\mu m$, \cii~157$\mu m$ with $\kappa(O)$ > 10. These lines are notorious for being formed in photodissociation regions that are predominantly neutral and have physical and chemical processes that are not taken into account in the current version of the MOCASSIN code and, therefore, are not expected to be reproduced. The next important discrepancy to note is that of the He~{\sc ii} $\lambda$4686 recombination line with $\kappa(O) \simeq 7$. The \heii~lines are important constraints for the temperature of the central ionising source. Fortunately, in this case we have more than one observation of the line \heii~$\lambda$5411, which we can use to help understand the discrepancy. As we can see in Table \ref{tab:linecomps}, the measurement of the line \heii~$\lambda$5411, which was obtained by \citet{MUSE3132}, is well reproduced by the model. The intensity value of the \heii~$\lambda$4686 line, which is not present in the MUSE data, was obtained by \citet{Tsamis2003}, who also measured \heii~$\lambda$5411 to be 0.004 (H$\beta$=1). Both measurements by \citet{Tsamis2003} are systematically lower than the values obtained by the model. More interestingly, the \heii~$\lambda$4686 line shows a large variation in the long-slit spectroscopy determinations found in the literature, as can be seen in Table \ref{tab:heii}, which indicates that slit position is an important factor. However, the upper limits in cases with multiple measurements are always considerably higher than the value obtained by \citet{Tsamis2003}. If we take the average of the literature values we get $I_{\text{He~{\sc ii}}4686} \approx 0.14 \pm 0.09$, which is in agreement with the value obtained by the model. Other lines that we use as constraints and were measured by \citet{Tsamis2003}, which show discrepancies, are \neiii~$\lambda$3968 and \oii~$\lambda$3727 and in both cases the model predicts higher intensities. The fact that for all these lines, which are in the bluer region of the spectrum, the model values are systematically higher than the measurements seems to indicate a relation to the extinction correction, since other lines measured by the same author in the redder parts of the spectrum agree well with the model.  

For the lines \oii~$\lambda$7332  and \siii~$\lambda$9072 we also see a relevant discrepancy; however, these lines can be strongly affected by telluric emission and, therefore, prone to higher uncertainties depending on how the sky subtractions have been performed on the data. For \oii~$\lambda$7332, which was measured by \citet{MUSE3132} and \citet{Tsamis2003} to be 0.036 and 0.056, respectively, there is a significant difference indicating that the uncertainty on this line may be underestimated. 

We also included in Table \ref{tab:linecomps} selected recombination emission line measurements from \citet{Tsamis2003}. Compared to the model, the oxygen and nitrogen lines show considerable discrepancies of $\kappa(O)>4$. The line from carbon shows a small discrepancy, although we believe that the uncertainties cited by the authors are considerably underestimated. A more conservative value of $\Delta O / O = 0.5$ for these lines would place the carbon line with $\kappa(O) = -0.879$, while the lines for oxygen and nitrogen would still be discrepant. Our model considers only uniform abundances throughout the main nebula and therefore cannot explain such a discrepancy. To investigate that in more detail we would need a two-phase model (with high and low metalicity zones) and better observational constraints, as these recombination lines tend to be very faint and may appear in regions of the spectrum where there is considerable overlap with other lines, making the task of measuring them precisely quite challenging.


\begin{table}
\centering
\caption{\heii 4686 Line Intensities obtained in the literature}
\begin{tabular}{|c|c|}
\hline
Literature Reference & Intensity ($H\beta$=1)\\ \hline
\citet{AF64} & 0.27 \\
\citet{TP77} & 0.0458 - 0.258 \\
\citet{1976ApJ...203..636K} & 0.24 \\
\citet{BDG90} & 0.053 - 0.092 \\
\citet{DissertacaoMonteiro2000} & 0.106 \\
\citet{Krabbe2006} & 0.0707 \\
\hline
\end{tabular}
\label{tab:heii}
\end{table}

\begin{table*}
\label{tab:linecomps}
\centering
\caption{Comparison of observed emission line intensities and diagnostic ratios to their respective model values. Lines used as constraints in the optimization process (see sec. \ref{sec:optimize}) are indicated by the $^*$ superscript. Also shown here are the adopted tolerance $\Delta O/O$ and resulting quality factor $\kappa(O)$ for each line and diagnostic ratio. }
\begin{tabular}{lcccccc}
\hline
\textbf{Line} & \textbf{$I_{\lambda}$} & \textbf{$\sigma_{I_{\lambda}}$} & \textbf{$\Delta O / O$} & \textbf{Model} & \textbf{$\kappa(O)$} & \textbf{Ref.}\\
\hline
\textbf{Optical CE Lines ({\AA})} & & & & &  & \\
\ha\ 6563 & 2.880 & 0.022 & 0.050 & 2.860 & -0.146 & 2 \\
H$\gamma$ 4341$^*$ & 0.410 & 0.021 & 0.150 & 0.468 & 0.952 & 1 \\
\hei\ 5877$^*$ & 0.165 & 0.006 & 0.102 & 0.170 & 0.313 & 2 \\
\hei\ 6678$^*$ & 0.046 & 0.008 & 0.524 & 0.048 & 0.110 & 2 \\
\hei\ 4471$^*$ & 0.060 & 0.003 & 0.150 & 0.062 & 0.193 & 1 \\
\heii\ 4686 & 0.040 & 0.002 & 0.150 & 0.100 & 6.574 & 1 \\
\heii\ 5411$^*$ & 0.007 & 0.011 & 4.397 & 0.008 & 0.024 & 2 \\
\niA\ 5200$^*$ & 0.082 & 0.041 & 1.494 & 0.172 & 0.817 & 2 \\
\nii\ 6549$^*$ & 1.796 & 0.014 & 0.050 & 1.885 & 0.984 & 2 \\
\nii\ 6585$^*$ & 5.563 & 0.017 & 0.050 & 5.556 & -0.027 & 2 \\
\nii\ 5756$^*$ & 0.076 & 0.006 & 0.240 & 0.086 & 0.540 & 2 \\
\oi\ 6302$^*$ & 0.335 & 0.014 & 0.300 & 0.412 & 0.788 & 2 \\
\oii\ 3727$^*$ & 5.510 & 0.276 & 0.150 & 6.587 & 1.278 & 1 \\
\oii\ 7322$^*$ & 0.067 & 0.020 & 0.898 & 0.101 & 0.637 & 2 \\
\oii\ 7332$^*$ & 0.035 & 0.008 & 0.651 & 0.120 & 2.435 & 2 \\
\oiii\ 5008$^*$ & 8.362 & 0.013 & 0.050 & 8.456 & 0.230 & 2 \\
\oiii\ 4364$^*$ & 0.040 & 0.006 & 0.458 & 0.053 & 0.770 & 1 \\
\neiii\ 3869 & 1.180 & 0.059 & 0.150 & 1.060 & -0.764 & 1 \\
\neiii\ 3968 & 0.510 & 0.025 & 0.150 & 0.319 & -3.347 & 1 \\
\sii\ 6732$^*$ & 0.507 & 0.011 & 0.100 & 0.463 & -0.962 & 2 \\
\sii\ 6718$^*$ & 0.518 & 0.016 & 0.100 & 0.468 & -1.068 & 2 \\
\siii\ 6312$^*$ & 0.022 & 0.010 & 1.299 & 0.035 & 0.527 & 2 \\
\siii\ 9072$^*$ & 0.384 & 0.055 & 0.428 & 0.632 & 1.397 & 2 \\
\cliii\ 5539 & 0.008 & 0.013 & 4.913 & 0.008 & 0.021 & 2 \\
\cliii\ 5519 & 0.010 & 0.012 & 3.660 & 0.010 & -0.012 & 2 \\
\ariii\ 7136$^*$ & 0.273 & 0.006 & 0.100 & 0.265 & -0.321 & 2 \\

\hline
\textbf{Recombination Lines ({\AA})} & & & & &  & \\
\textsc{C\,ii}~4267 & 0.0070 & 0.0007 & 0.30000 & 0.0049 & -1.359 & 1  \\
\textsc{N\,ii} 4630 & 0.0004 & 0.0001 & 0.30000 & 0.00003 & -9.873 & 1  \\
\textsc{N\,ii} 5678 & 0.0008 & 0.0001 & 0.30000 & 0.00022 & -4.921 &  1 \\
\textsc{O\,ii} 4069 & 0.0041 & 0.0004 & 0.30000 & 0.00123 & -4.589 &  1 \\

\hline
\textbf{Infrared Lines ($\mu m$)} & & & & &  & \\
\nii~122 & 0.041 & 0.004 & 0.300 & 0.088 & 2.863 & 3 \\
\niii~57$^*$ & 0.675 & 0.068 & 0.300 & 0.823 & 0.752 & 3 \\
\oiii~52$^*$ & 1.699 & 0.170 & 0.300 & 1.942 & 0.508 & 3 \\
\oiii~88$^*$ & 1.198 & 0.120 & 0.300 & 0.870 & -1.220 & 3 \\
\oi~63 & 0.446 & 0.045 & 0.300 & 0.006 & -16.533 & 3 \\
\oi~146 & 0.013 & 0.001 & 0.305 & 0.001 & -11.586 & 3 \\
\cii~157 & 0.056 & 0.006 & 0.303 & 0.173 & 4.309 & 3 \\

\hline
\textbf{Ultraviolet Lines ({\AA})} & & & & &  & \\
\heii\ 1640$^*$ & 0.851 & 0.113 & 0.398 & 0.649 & -0.809 & this work \\
\textsc{C\,iii]}1906+1908$^*$ & 2.642 & 0.135 & 0.153 & 2.354 & -0.810 & this work \\

\hline
\textbf{Line Diagnostics} & & & & &  & \\
\sii\ 6731/6717$^*$ & 0.980 & 0.037 & 0.113 & 0.989 & 0.095 &   \\
\oiii\ (4959+5007)/4363$^*$ & 278.733 & 42.509 & 0.458 & 210.939 & -0.740 &  \\
\nii\ (6584+6546)/5754$^*$ & 97.213 & 7.777 & 0.240 & 86.439 & -0.546 &  \\
\siii\ (9069+9531)/6312$^*$ & 59.642 & 27.194 & 1.368 & 63.294 & 0.069 &  \\

\textbf{Absolute Flux} & & & & &  & \\
F(H$\beta$) & $9.18\times 10^{-11}$ & $4.59\times 10^{-12}$ & 0.050 & $9.23\times 10^{-11}$ & 0.111 & \\
(erg~cm$^{-2}$~s$^{-1}$) & & & & &  & \\
\hline
\multicolumn{7}{l}{\tiny 1 - \citet{Tsamis2003}, 2 - \citet{MUSE3132}, 3 - \citet{Liu2001}} \end{tabular}
\end{table*}

Another important constraint for the model is the ionic stratification for different species. This is verified through the ability of the model to reproduce different emission line structures and dimensions produced by different ions and ionisation stages. In Fig. \ref{fig:MUSEcomp} we show a comparison of model emission line maps and MUSE maps obtained for characteristic lines of low, middle, and high ionisation stages of the elements H, He, O and N. In Fig. \ref{fig:profilecomp} we show a comparison of the \hb~MUSE image to the model, overlaid with contours showing that the model reproduces well not only the dimensions but also the observed stratification. This is also seen in the right panel of the latter figure, where cuts along the E-W direction passing through the central star position show how well the model reproduces the relative intensity variations of the lines \hb~as well as \nii. This is a particularly important constraint, as it relates to the total gas and dust mass available to be ionized and the central star capacity to do so. In our case, it is a very stringent constraint, as the physical size of the density structure is set by the very precise Gaia distance which is in the range 736 < D < 769~pc, as discussed by \citet{2022NatAs...6.1421D} and used here as the distance of the nebula. The fact that the model reproduces the ionisation stratification details and sizes in distinct lines with such good precision is remarkable and is evidence that the homologous expansion assumption is a reasonable one.

\begin{figure*}
	\includegraphics[width=\textwidth]{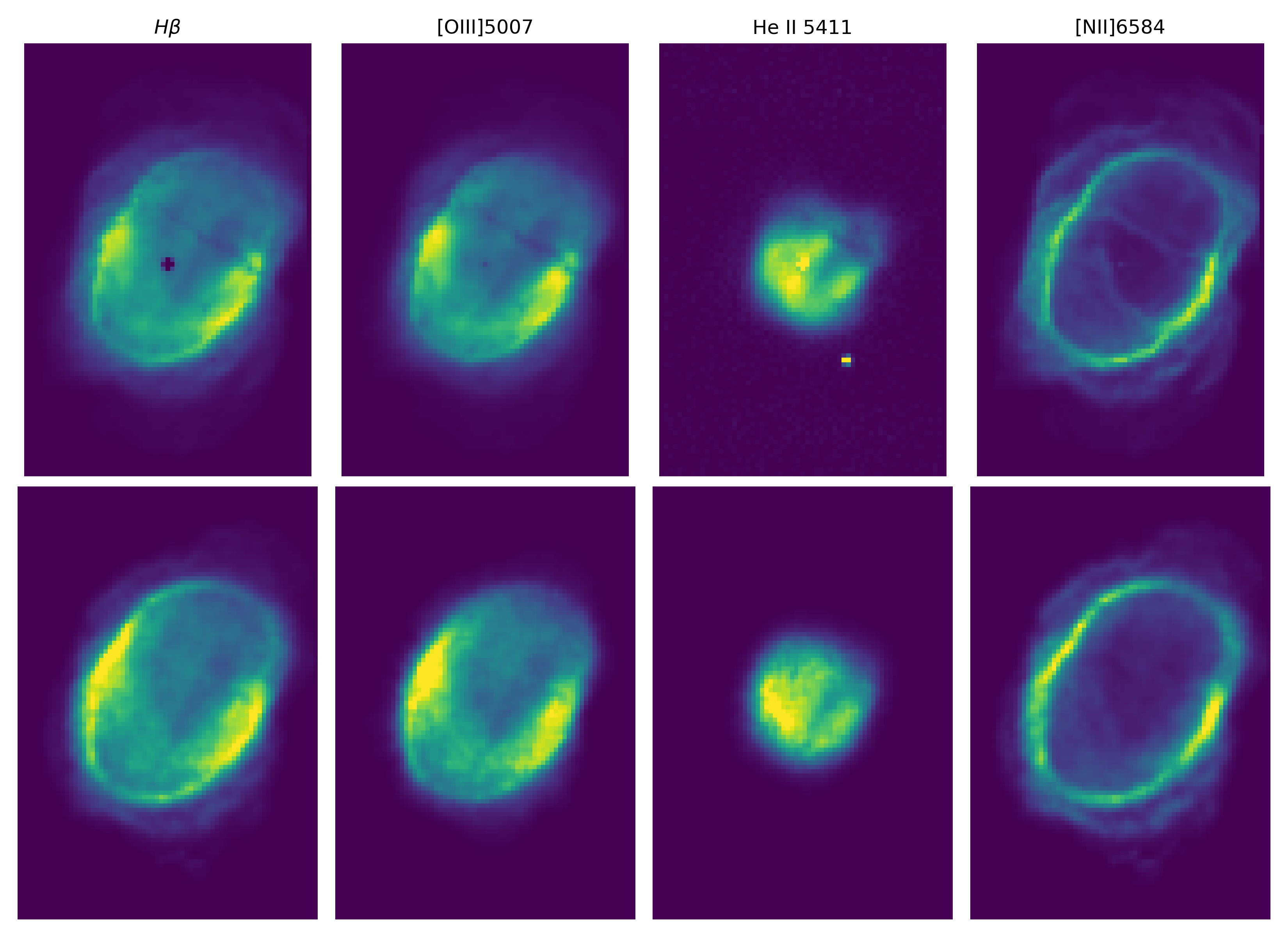}
    \caption{Comparison of model emission line intensity in $ergs/cm^2/s$ (lower row) to MUSE maps (upper row) obtained for characteristic lines of low, middle and high ionization stages of the elements H, He, O and N. Both MUSE and model images are displayed in linear scale.}
    \label{fig:MUSEcomp}
\end{figure*}

\begin{figure*}
	\includegraphics[scale=0.6]{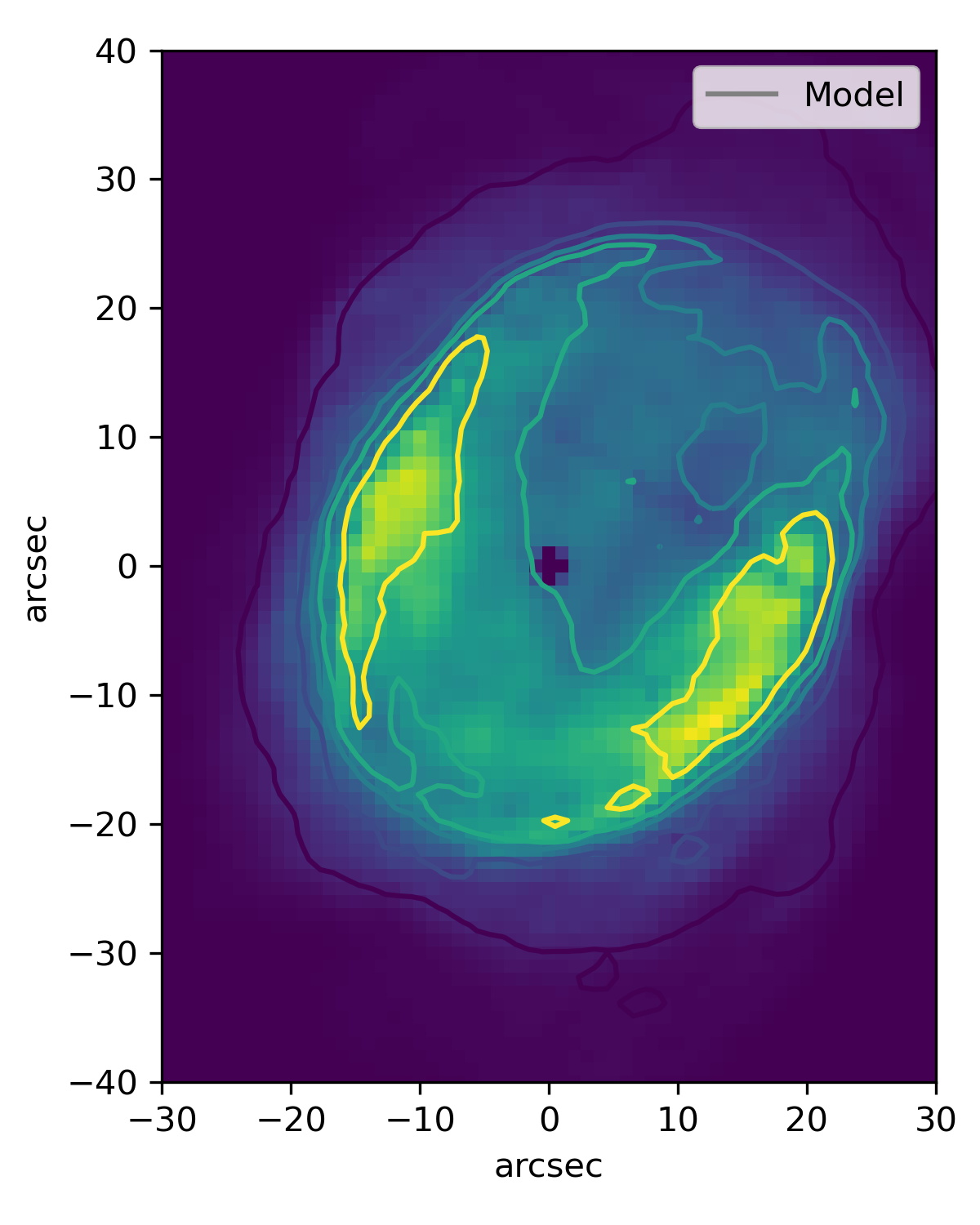}
	\includegraphics[scale=0.67]{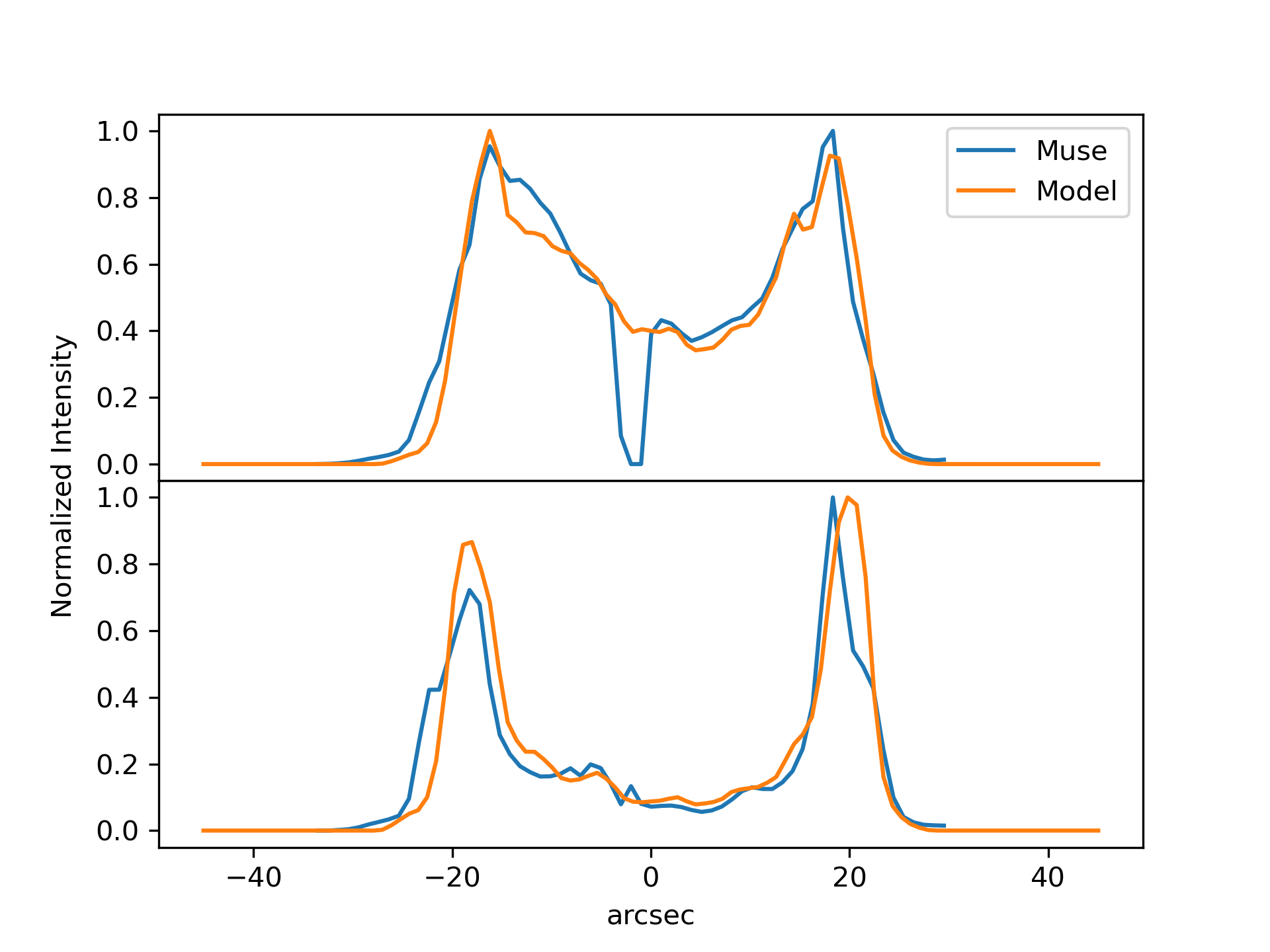}
    \caption{Comparison of observed \ha~emission line image obtained from MUSE (left panel) to the model (overlaid contour) as well as E-W relative intensity profiles from \ha~and \nii~(right panel) }
    \label{fig:profilecomp}
\end{figure*}

\begin{figure*}
	\includegraphics[width=\textwidth]{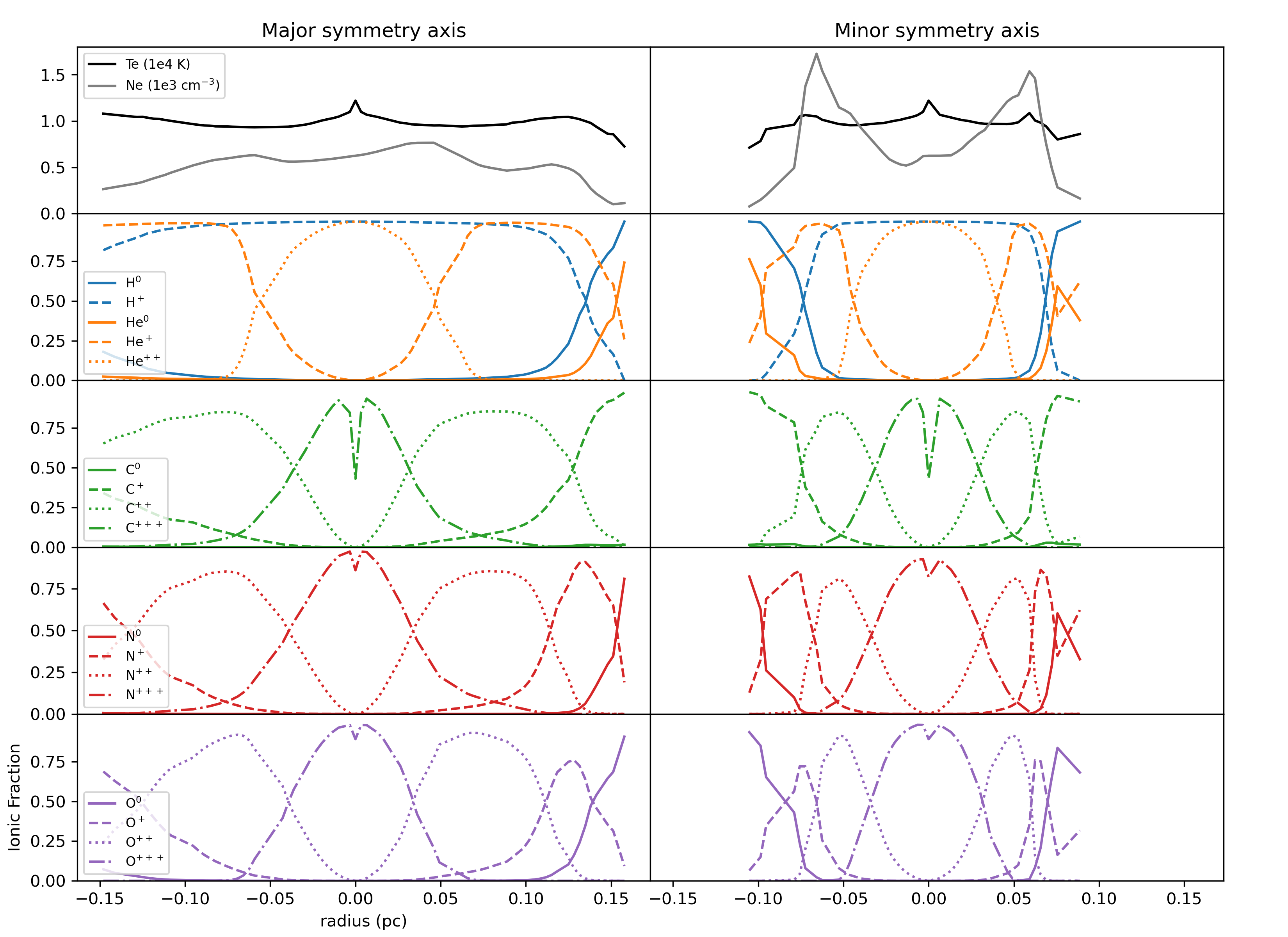}
    \caption{Electron temperature, electron density and ionic fraction profiles for two directions through the model nebula. Left: along the x-axis; right: along z-axis, both through the nebula centre.}
    \label{fig:ionic_profiles}
\end{figure*}

The ionisation structure of the nebula is complex, but we can look at characteristic directions to gain a better understanding of the conditions. In the model, we find that 6\% of the cells where $N_e>0$ have H$^+$/H > 99\% and 30\% have H$^+$/H < 1\%, with 69\% of cells with values between. In Fig. \ref{fig:ionic_profiles} shows the electron temperature, electron density and ionic fraction profiles for two directions through the model nebula (Left: along the x-axis; right: along z-axis, both through the nebula centre).


\begin{figure}
	\includegraphics[scale=0.41]{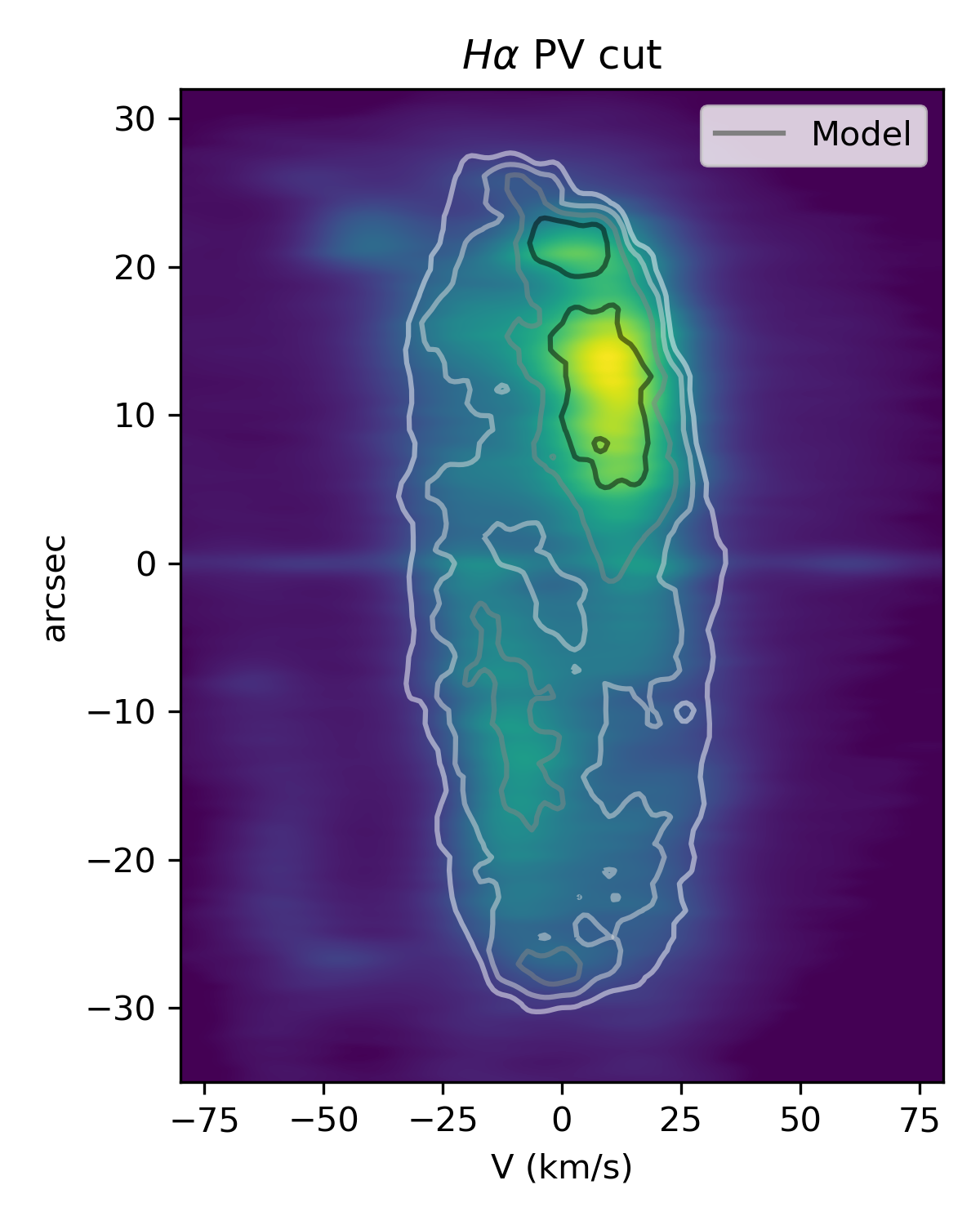}
	\includegraphics[scale=0.41]{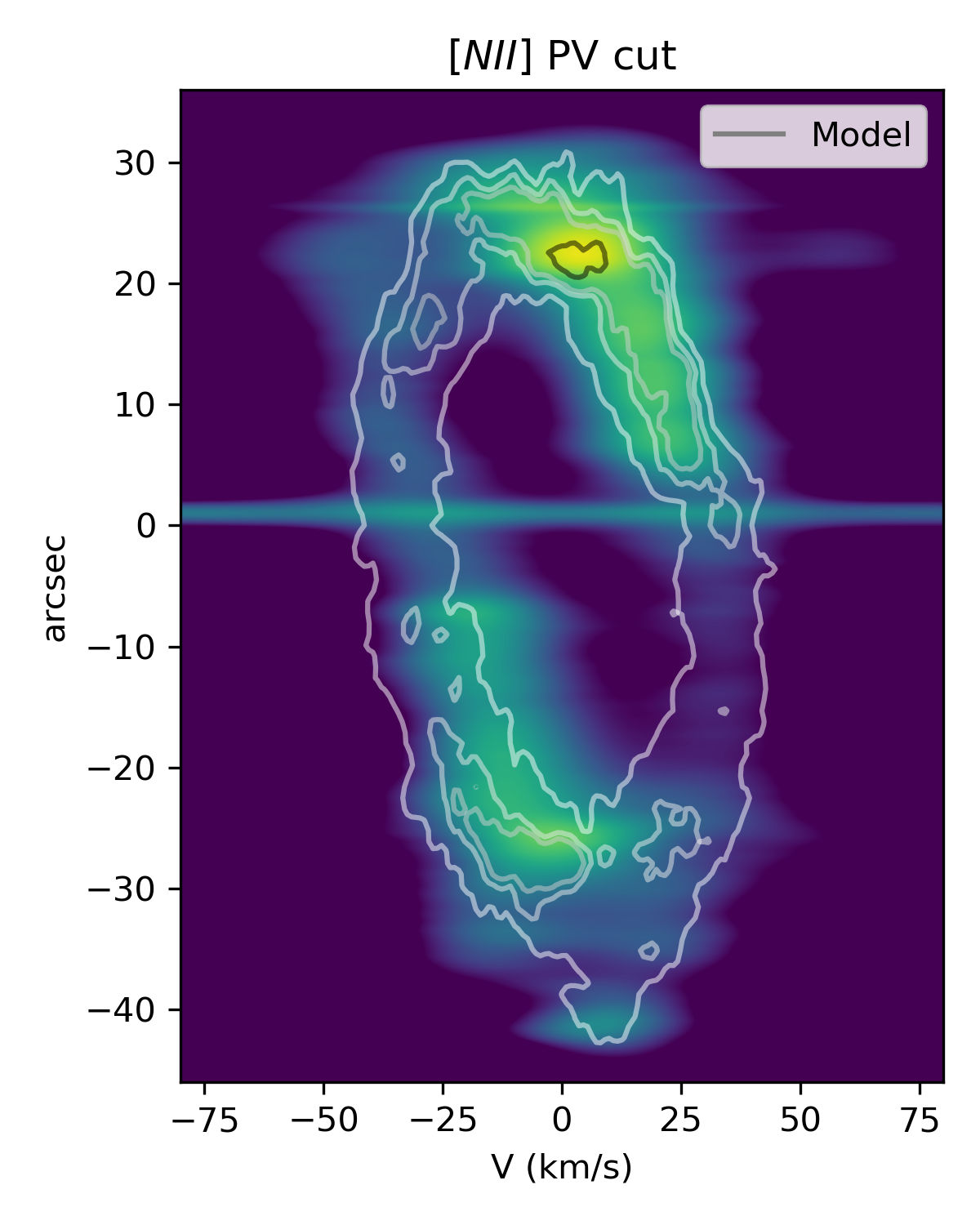}
    \caption{Position velocity diagrams for cuts along the major axis of the nebula for the \ha~(left panel) and \nii~(right panel) images compared to their respective results from the model (overlaid contours).}
    \label{fig:PVmodcomp}
\end{figure}

The velocity field from the homologous expansion assumption allows us to perform another sanity check of the model. With the velocity field combined with the emissivities obtained by the model, we can simulate the observed PV diagrams for any given emission line. In Fig. \ref{fig:PVmodcomp}, we show the results for cuts along the major axis of the nebula for the \ha~and \nii~emission lines, which were observed by the SAM-FP, as described in sec. \ref{sec:samfp}. The model PV diagram contours are overlaid on the image obtained from the data. There is good agreement for the overall structure of the PV diagram with some evident discrepancies in the fainter outer parts of the nebula, where the homologous expansion assumption probably starts to be a poor approximation. The model also does not well reproduce the inner regions due to the resolution limit of the observational data, especially in \ha~, which is mainly produced in the lower velocity regions of the nebula.

A detailed comparison to the spatially resolved data from MUSE and a discussion about the empirical abundance determination and how it compares to the results of the model in this work will be made in a forthcoming paper by Bouvis et al. in preparation. The spatial distribution of electron density and temperature derived from different pseudo-slit, as well as along radial directions from the MUSE observations and MOCASSIN model have been thoroughly examined and compared employing the SATELLITE code \citep{Akras2022}. These comparisons were useful to improve the current photoionisation model of NGC~3132.

We also compared the model escaped spectral energy distribution (SED) with the available observations. In Fig. \ref{fig:SED} we present the Escaped SED obtained by the model compared to the available observed data. Infrared photometry from IRAS, WISE and AKARI obtained from Vizier\footnote{\url{https://vizier.cds.unistra.fr/viz-bin/VizieR}} and infrared spectra from ISO from \citet{Liu2001} and Spitzer from \citet{Delgado2014} are also plotted. The Spitzer spectra were taken using slits that do not cover the entire nebula, so it was scaled to correspond to the ISO spectrum and to be consistent with the available photometry. Also shown is the SED of the ionising source used and a black-body curve of the same temperature for comparison, as well as the measured photometry for the central source obtained by \citet{2022NatAs...6.1421D} and \citet{2023ApJ...943..110S}.  For the dust in the outer regions, we explored the dust compositions of SiC, graphite, and silicate, as shown in the lower panel of Fig. \ref{fig:SED}. It can be seen from the figure that there is reasonable agreement of the dust continuum of the escaped SED with the available photometry for all three grain types investigated. 

The dust composition of SiC gave slightly better results around the $10\micron$ compared to the other grain types, so this was the dust type adopted in the final model fit and abundance optimisation. The dust mass obtained with SiC dust grains in the size range of $0.012\micron$ to $0.7\micron$, in the usual power law distribution of sizes with $dn/da \propto a^{-3.5}$, was of $2.6\times10^{-3}M_{\odot}$. The temperatures obtained with the dust distribution and composition were in the range of 40K to 80K. It should be noted that SiC dust grains are expected to be found in C-rich stars (C/O > 1) according to \citet{2007ApJ...671.1669S}, which is consistent with the composition we find for the nebula.

Overall, the other dust grain types require about the same dust mass to reproduce the observations in the dust continuum, but with smaller grain sizes. However, with our current dust distribution, to reach the observed emission around $10\micron$, distributions with smaller grains reaching the inner regions of the nebula would be required for graphite and silicate dust. In such cases the grains interfere considerably with the heating and ionisation structure of the gas phase the nebula, leading to a worse reproduction of the observed emission-line fluxes and a poorer model fit overall. It is possible that a combination of grain types and size distributions may lead to a better fit of the data: indeed, SiC dust grains are mixed with graphite/amorphous carbon in C-rich AGB stars \citep{2007MNRAS.376..313G} but this is beyond our scope.

For the dust surrounding the central source, large grain sizes ($0.2\micron$ to $1.0\micron$) were required to reproduce the observations.  It suggests that a binary trapped central structure \citep{2022NatAs...6.1421D}, such as a disk, may be a reservoir of large dust grains. In a simple AGB outflow, the gas expands adiabatically and hence the density drops rapidly. The collisions between particles (both gas and dust) would be substantially reduced in a short space of time. On the other hand, a disk is gravitationally bound by the central star to some extent and could sustain reasonably high densities for a long time. That would allow dust grains to stick together and coagulate. This coagulation to large dust grains may be commonly found in disks, not only in proto-planetary disks \citep{2017A&A...605A..16F}, but also in PNe.

\begin{figure}
	\includegraphics[width=\columnwidth]{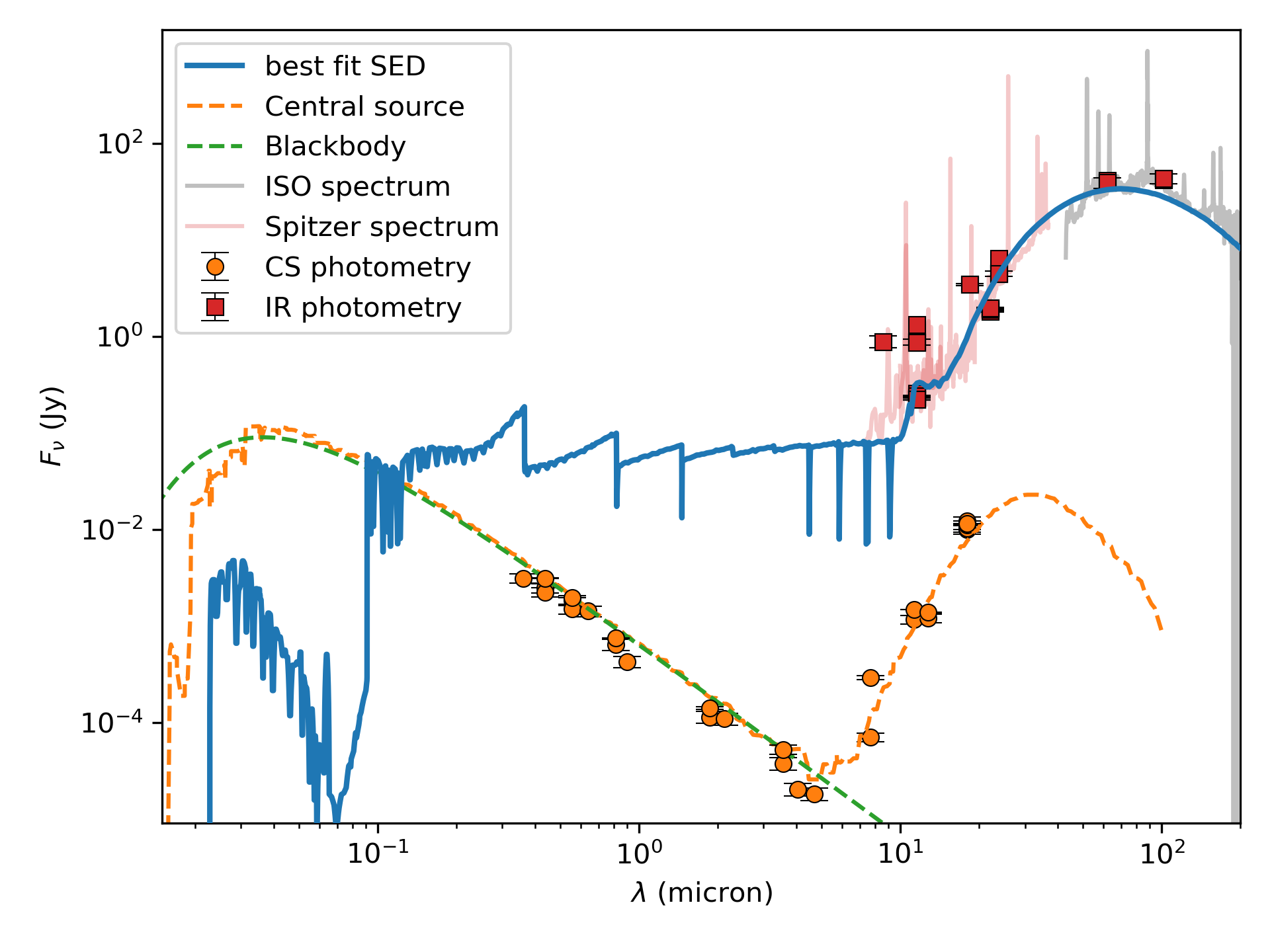}
	\includegraphics[width=\columnwidth]{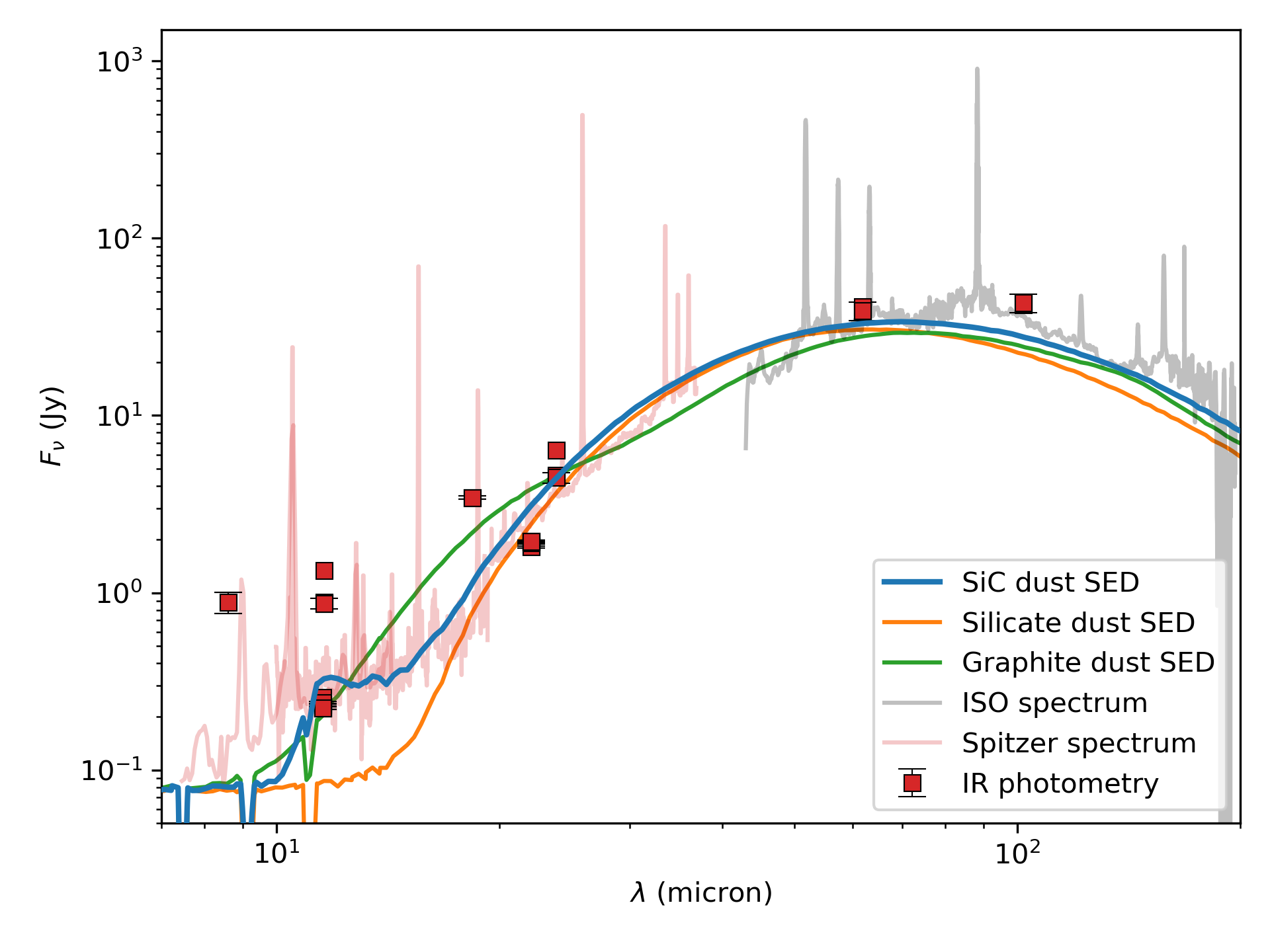}
    \caption[]{ Escaped SED obtained by the best fit model (blue line) compared with the available observed data. Infrared photometry from IRAS, WISE and AKARI (red squares) was obtained from Vizier. Infrared spectra from ISO from \citet{Liu2001} and Spitzer from \citet{Delgado2014} are also plotted (light red and gray lines respectively). Also shown is the used ionizing source SED (orange line) and a blackbody curve of the same temperature (dashed green line) for comparison with the measured photometry for the central source obtained by \citet{2022NatAs...6.1421D} and \citet{2023ApJ...943..110S}. Lower panel shows details of the far infrared part of the spectrum and the model results using SiC (blue line), Silicate (orange line) and graphite (green line). }
    \label{fig:SED}
\end{figure}


\begin{table}
    \centering
    \caption{Summary of best fit NGC 3132 photoionization model parameters.}
    \label{tab:params_modelA}
    
    \resizebox{\columnwidth}{!}{%
    \begin{tabular}{lc}
    \hline
    Parameter & Value \\
    \hline
    \textbf{Ionizing Source:} & \\
    Effective Temperature ($T_\textrm{eff})$ & 140.0~kK \\
    Luminosity & $(212 \pm 11)L_\odot$ \\
    $M_i$ & $(2.7 \pm 0.2)M_{\odot}$ \\
    $M_\text{core}$ & $(0.64 \pm 0.04)M_{\odot}$ \\
    
    {} & {} \\
    \textbf{Central Source Shell:} & \\
    Geometry & Spherical \\
    Inner, outer radius (cm) & $1.0 \times 10^{15}$, $5.0 \times 10^{15}$ \\
    Gas mass & $2.42 \times 10^{-5}\,\textrm{M}_{\odot}$ \\ 
    Dust mass & $2.42 \times 10^{-10}\,\textrm{M}_{\odot}$ \\ 
    He mass fraction ($X_\text{He}$) & $6.5 \times 10^{-3}$ \\ 
    C mass fraction ($X_\text{C}$) & 0.474 \\ 
    N mass fraction ($X_\text{N}$) & $4.7 \times 10^{-7}$ \\ 
    O mass fraction ($X_\text{O}$) & 0.466 \\ 
    Dust grain composition & Graphite \\ 
    Dust grain size range & $0.2\micron$ to $1.0\micron$ \\
    Dust grain size distribution & $dn/da \propto a^{-3.5}$ \\

    {} & {} \\
    \textbf{Main Nebular Structure:} & \\
    Geometry & derived from velocity field \\
    $n_{H}$ interval & from about 50 to 1900  (Fig.~\ref{fig:strutcuts}) \\
    size [cm] & $5.074 \times 10^{17}$ \\
    Gas Mass [M$_\odot$] & 0.14 \\
    Dust mass & $1.0\times10^{-2}M_{\odot}$ \\ 
    Dust grain composition & 100\% SiC dust \\ 
    Dust grain size range & $0.012\micron$ to $0.7\micron$ \\
    Dust grain size distribution & $dn/da \propto a^{-3.5}$ \\
    
    {} & {} \\
    \textbf{Gas Abundances:} & \\

    He/H & 0.117$^{+0.01}_{-0.01}$ \\[3mm]
    C/H [$10^{-3}$]& $1.17^{+0.05}_{-0.07}$ \\[3mm]
    N/H [$10^{-4}$] & $2.27^{+0.3}_{-0.2}$ \\[3mm]
    O/H [$10^{-4}$] & $5.80^{+0.7}_{-0.6}$ \\[3mm]
    S/H [$10^{-5}$] & $1.06^{+0.2}_{-0.2}$ \\[3mm]
    Ar/H [$10^{-6}$] & $3.10^{+0.5}_{-0.5}$ \\[3mm]
    Ne/H [$10^{-4}$] & $1.20^{+0.2}_{-0.2}$ \\[3mm]
    Cl/H [$10^{-7}$] & $1.70^{+0.2}_{-0.2}$ \\[3mm]
    
    \hline
    \end{tabular}%
    }
\end{table}


Finally, we look at the results for the abundances of the elements determined in our model fitting.  The results presented in Table \ref{tab:pn-abundances} show a comparison between the elemental abundances derived from the 3D photoionization model and those reported in the literature. The helium abundance from the 3D model is consistent with values from the literature, showing only minor deviations that may arise from differences in methodologies. The carbon abundance derived from the model is significantly higher than the values found in the literature, in \citet{Tsamis2003} and \citet{MUSE3132}. The lower value reported by \citet{Tsamis2003} is likely due to the use of collisionally excited lines (CELs) observed in IUE apertures that do not encompass the entire nebula, requiring aperture corrections. The fact that in \cite{Tsamis2004}, the same authors obtain $1.27\times10^{-3}$ for the abundance of carbon obtained from recombination lines, which is better in agreement with our model value, together with the more detailed scaling we have performed for the IUE data, as described in \ref{sec:iuedata}, indicates that the aperture correction may indeed be the cause. The Nitrogen abundance is comparable with the values derived by \citet{Tsamis2003} and \citet{Krabbe2006}, but is lower than that found in \citet{MUSE3132}, probably due to their use of a simplified one-dimensional model. 

The abundance of oxygen is at the lower end of the literature values, which range from $5.72\times10^{-4}$ to $8.60\times10^{-4}$ and the abundances of neon, sulphur, chlorine and argon are consistently lower than values from the literature. This discrepancy could be related to differences in methodologies and assumptions made, such as ionisation correction factors (ICF) for these elements and the fact that our model gives slightly higher electron temperatures than those obtained from observations.

The value derived from the 3D model for the carbon-to-oxygen ratio (C/O) is significantly higher, while the nitrogen-to-oxygen ratio (N/O) is in agreement with the literature values derived from collisionally excited lines. Our C/O value is also higher than 0.81, ratio obtained by \citet{Tsamis2004} from recombination lines. The lower C/O values from the literature would indicate a progenitor star of about 1$M_{\odot}$ according to \citet{2018MNRAS.475.2282V}, which is in disagreement with the recent results of \citet{2022NatAs...6.1421D}. 

In fact, we have very stringent constraints on the progenitor star from the model results. The final properties of the ionising source, as shown in Table \ref{tab:params_modelA} and illustrated in Fig. \ref{fig:HRD}, indicate that the progenitor star has a mass of $(2.7 \pm 0.2)M_{\odot}$. The core mass of $(0.64 \pm 0.04)M_{\odot}$ and the progenitor mass were obtained by interpolating the grids of the intermediate-mass star evolutionary models of \citet{Bertolami2016} (shown in the figure as dashed blue lines). In each cooling track, we show the corresponding intervals of dynamical age $(2061 \pm 412)yr$ determined from the homologous velocity field fitted. In the zoomed-in panel in Fig. \ref{fig:HRD} we can see the interpolated cooling track (orange dashed line) as well as the corresponding dynamical age interval showing the agreement of the ionising source luminosity and $T_{eff}$ properties derived from our model fit with the predictions of the \citet{Bertolami2016} models.

\begin{table*}
\caption{Abundances from the 3D model compared with values from the literature. Abundances are ratio of elements to hydrogen by number. The uncertainties of the model abundances have been approximated by symmetric values for clarity.
}
\begin{tabular}{lcccc}
\hline
                           & 3D Model fit        & Tsamis (2003)      & Krabbe (2003)         & Monreal-Ibero (2020) \\
\hline
{\bf He/H}                       & 0.117 $\pm$ 0.010   & 0.12               & 0.126 $\pm$ 0.007     & 0.124                \\
{\bf C/H} [$10^{-4}$]     & $11.7 \pm 0.7$      & $3.20$~$^{(a)}$    & --                    & $7.10$  \\
{\bf N/H} [$10^{-4}$]     & $2.27 \pm 0.32$   & $2.40$             & $2.91 \pm 0.23$     & $3.60$  \\
{\bf O/H} [$10^{-4}$]     & $5.80 \pm 0.71$   & $6.60$             & $5.72 \pm 0.35$     & $8.60$  \\
{\bf Ne/H} [$10^{-4}$]    & $1.20 \pm 0.20$   & $3.10$             & $3.11 \pm 0.20$     & $3.50$  \\
{\bf S/H} [$10^{-5}$]     & $1.06 \pm 0.20$   & $1.10$             & $1.14 \pm 0.05$     & $1.30$  \\
{\bf Cl/H} [$10^{-7}$]    & $1.70 \pm 0.20$   & $2.30$             & $2.61 \pm 0.19$     & $2.50$  \\
{\bf Ar/H} [$10^{-6}$]    & $3.10 \pm 0.50$   & $5.40$             & --                  & $3.80$  \\
{\bf Method}              &  3D phot. model   & scanning long-slit &  long-slit          & 1D phot. model       \\
                    &                   &                    &                     &                      \\
\hline
{\bf C/O}                 & $2.02 \pm 0.28$   & 0.48               & --                  & 0.83                 \\
{\bf N/O}                 & $0.39 \pm 0.38$   & 0.36               & 0.51                & 0.42                 \\
\hline
\multicolumn{5}{l}{\tiny $^{(a)}$Carbon abundance is from collisionally excited lines. The authors obtain $1.27\times10^{-3}$ from recombination lines.} \\
\label{tab:pn-abundances}
\end{tabular}
\end{table*}


\begin{figure}
	\includegraphics[width=\columnwidth]{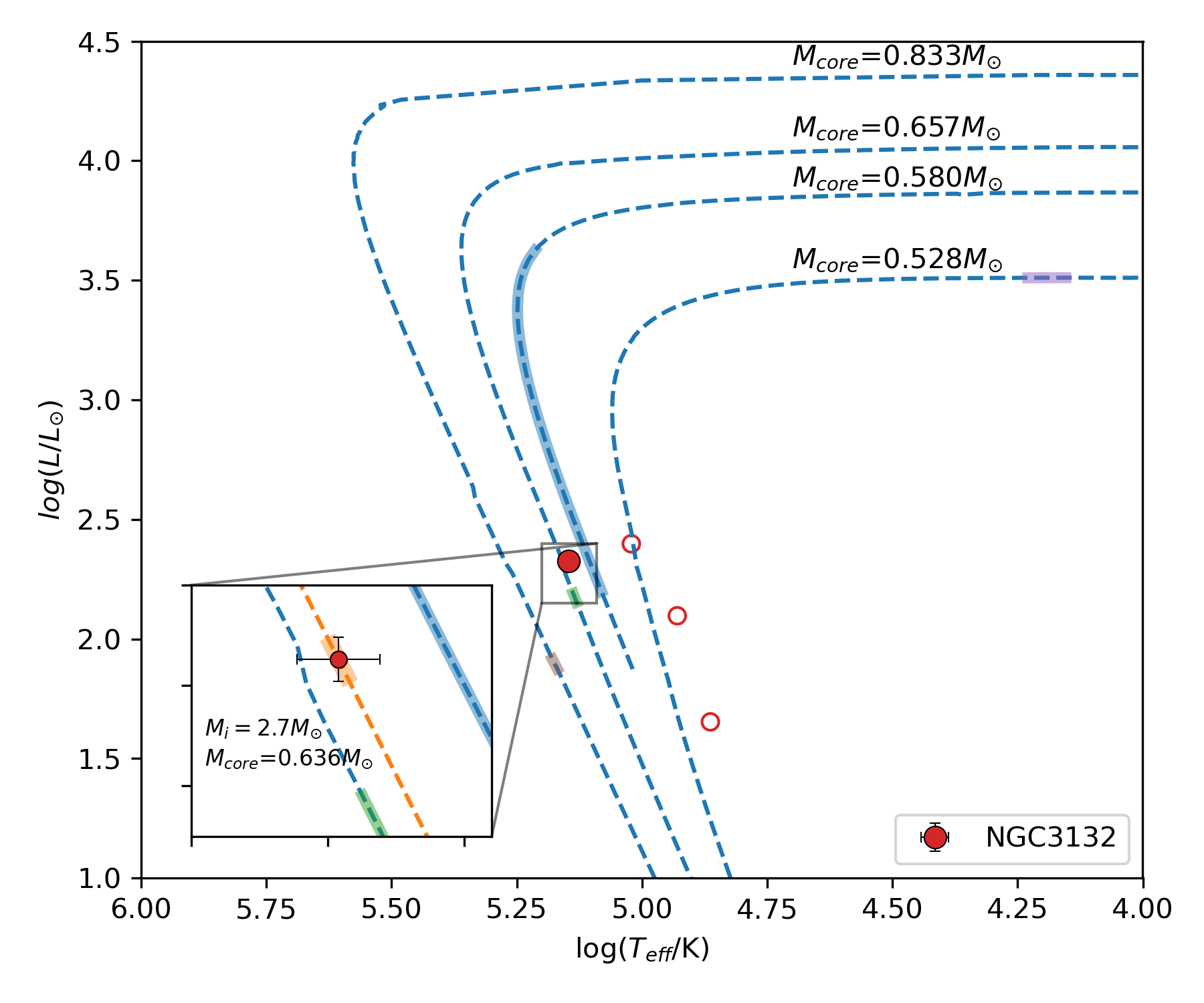}
    \caption[]{HR diagram showing the central source $T_{eff}$ and luminosity, as derived from the photoionization model fit interpolated in the grids of the intermediate-mass star evolutionary models of \citet{Bertolami2016} (shown in the figure as dashed blue lines), to obtain core and progenitor star masses. In each cooling track, the corresponding intervals of dynamical age $(2061 \pm 412)yr$, determined from the homologous velocity field fitted, are shown (coloured thick lines). In the zoomed-in panel the interpolated cooling track (orange dashed line) as well as the corresponding dynamical age interval are shown.}
    \label{fig:HRD}
\end{figure}

\section{Conclusions}
\label{sec:conclusion}

In this study, we presented a detailed high-definition photoionisation model for the nebula NGC 3132. The methodology incorporated the use of a global optimisation technique with the MOCASSIN code to derive the chemical abundances of the most important elements such as C, N and O for. By incorporating a robust variety of observational constraints, including spatially resolved spectroscopic data, velocity-resolved line profiles, emission maps, and photometry from recent high-quality observational data from instruments such as MUSE from the VLT and MIRI and NIRCam, we achieved consistent reproduction of the observed nebular properties. 

In this process, the high-resolution spectroscopic data obtained from the SAM-FP instrument and the detailed kinematic structure derived from it confirm the existence of a complex, expanding gas structure with signatures of substructures and bipolar cavities, consistent with previous studies. We incorporated the spatially resolved velocity data from the SAM-FP data cubes to derive detailed, three-dimensional gas and dust density structures based on the assumption of homologous expansion. The resultant structure, which reproduces well the available spatially resolved observational constraints, reveals the complexity of the nebula, including multiple ring-like structures and a low-density cavity. 

To better reproduce the observations, we also had to include a combined dust and gas shell surrounding the central ionising source, in agreement with what was detected in \citet{2022NatAs...6.1421D}. Although initial models using standard blackbody sources or NLTE stellar atmospheres provided a basic fit, they failed to match the diagnostic temperature constraints, consistently predicting hotter plasmas. The incorporation of a dust and gas shell, with inner and outer radii (67 and 334 AU, respectively), a gas mass of $2.42 \times 10^{-5}\textrm{M}_{\odot}$, and a graphite dust grain mass of $2.42 \times 10^{-10}\textrm{M}_{\odot}$, constrained by the JWST photometric data, allowed better agreement with the observational data. The derived He-poor and C- and O-rich composition of the gas in the shell is consistent with stellar evolution models for a 3M$_{\odot}$ progenitor star. 

The optimisation approach allowed for an unbiased systematic exploration of the parameter space, yielding abundances that not only best reproduce the observed emission line strengths but also provide uncertainties. The results of the model with the optimised elemental abundances for He, C, N, O, and S show good agreement with the available observational data. The model successfully reproduces most emission line intensities within the adopted tolerances, with significant discrepancies for specific lines being explained by observational uncertainties such as sky removal and for lines such as \oi~63$\mu m$, \oi~146$\mu m$, and \cii~157$\mu m$, processes not currently accounted for in the model, such as photodissociation regions. The diagnostic ratios for the electron temperature are within the tolerance levels adopted but are close to the acceptable range limit, showing that there is still room for improvement. Two factors that may be affecting this are the detailed dust composition and the possible presence of a higher metallicity component in the main nebula, both of which should be explored in future work.

One notable exception are the intensities of the recombination lines of N and O. Unfortunately, we do not have good quality spatially resolved observations with high signal-to-noise to draw any reliable conclusions. In the case of C, the results of our model give reasonable collisional and recombination line intensities with a uniform abundance, in contrast to the result of \cite{Tsamis2004}, which found an abundance discrepancy factor of 4.4. The difference is likely due to the way we treat the IUE observations as described in sec. \ref{sec:iuedata}, taking into account details of the IUE aperture with spatially resolved data and emission line images of the model for the relevant transitions. Given the available observations, we cannot rule out the possibility of the existence of high metalicity phase, either as higher density knots or in some other form, existing within the main nebula as explored in the case of MGC~6153 done by \citet{2011MNRAS.411.1035Y} and \citet{2024A&A...689A.228G}, where authors find strong evidence confirming the existence of such a phase. It is possible that allowing for the presence of such a component would lead to a better agreement between model and observations in the case of the recombination lines as well as the derived electron temperature in the case of NGC~3132. However, to be able to make any inference on the subject, better and deeper spatially resolved observations of recombination lines intensities are needed.

The final fitted model presented here successfully reproduces all key observational constraints available for NGC~3132, particularly in terms of ionic stratification and detailed emission line structures across different ionisation stages. The good agreement between the model and the MUSE emission line maps demonstrates the ability of the model to accurately explain the ionisation complexity within the nebula. The precise distance measurement from Gaia played a critical role in constraining the model's physical size, further supporting the validity of the homologous expansion assumption. The simulated PV diagrams show good overall agreement with the observational data, although discrepancies in the fainter outer regions highlight the limitations of the expansion assumption at these distances. 

The results of the elemental abundances derived from the 3D photoionisation model provide insights into the stellar evolution of the progenitor star, specifically in the case of the carbon-to-oxygen (C/O) ratio, which is notably higher in our model, compared to previous determinations present in the literature. The discrepancy between our derived C/O ratio and the lower values reported in earlier studies points to a difference in the mass of the progenitor star. Stellar models predict that stars with initial masses around $2.7 \, M_{\odot}$, as we have found, which is consistent with the findings of \citet{2022NatAs...6.1421D}, typically produce nebulae with a higher C/O ratio, as they experience more efficient carbon synthesis during their evolution. In contrast, lower-mass stars (around $1 \, M_{\odot}$) tend to have a lower C/O ratio due to less extensive carbon production. Moreover, the properties of the ionising source, such as luminosity and $T_{\mathrm{eff}}$ agree with predictions from stellar evolution models, such as those in \citet{Bertolami2016}. Although we did not perform a detailed study of possible dust compositions and grain size distributions, the escaped SED of the model with dust composed of SiC grains, which provides a marginally better fit than that of graphite or silicate dust types, is inline with the C-rich environment scenario. 

This work underscores the effectiveness of combining optimisation-driven three-dimensional modelling with a wide range of observational constraints to derive robust and detailed physical and chemical properties of planetary nebulae. This work also demonstrates the potential of these methods to improve our understanding of stellar evolution, the chemical enrichment of the ISM, and the processes that govern gas and dust formation and evolution in these environments.

\section*{Data availability}

The SAMFP SOAR data used in this work are available upon request to the corresponding author. Some literature data used as constraints to the model were obtained from public archives such as Barbara A. Mikulski Archive for Space Telescopes (MAST; \url{https://mast.stsci.edu}), Combined Atlas of Sources with Spitzer IRS Spectra (CASSIS; \url{https://cassis.sirtf.com/atlas/welcome.shtml}), The Infrared Space Observatory (ISO; \url{https://nida.esac.esa.int/nida-cl-web/}).

Other data products are available upon request to the corresponding author. 

\section*{Acknowledgements}

This work has made use of the computing facilities available at the Laboratory of Computational Astrophysics of the Universidade Federal de Itajub\'{a} (LAC-UNIFEI). The LAC-UNIFEI is maintained with grants from CAPES, CNPq and FAPEMIG. KB and SA acknowledge support by the MINOTAUR research project implemented in the framework of H.F.R.I call ``Basic research financing (Horizontal support of all Sciences)'' under the National Recovery and Resilience Plan ``Greece 2.0'' funded by the European Union-NextGenerationEU (H.F.R.I. Project Number: 15665). R.W. and M.M. acknowledge support from the STFC Consolidated grant (ST/W000830/1).

This research has made use of the VizieR catalogue access tool, CDS, Strasbourg, France (DOI: 10.26093/cds/vizier). The original description of the VizieR service was published in A\&AS 143, 23. This research has made use of NASA's Astrophysics Data System. 

For the purpose of open access, the author has applied a CC BY public copyright licence (where permitted by UKRI, ‘Open Government Licence’ or ‘CC BY-ND public copyright licence’ may be stated instead) to any Author Accepted Manuscript version arising.

We also thank the referee for constructive comments and suggestions that have helped us improve the quality and clarity of our work.



\bibliographystyle{mnras}
\bibliography{refs} 




\bsp	
\label{lastpage}
\end{document}